\documentclass{scaleai-paper}

\usepackage{geometry}

\usepackage[square,sort&compress,numbers]{natbib}

\usepackage[utf8]{inputenc} 
\usepackage[T1]{fontenc}    
\usepackage{hyperref}       
\usepackage{url}            
\usepackage{booktabs}       
\usepackage{amsfonts}       
\usepackage{nicefrac}       
\usepackage{microtype}      
\usepackage{xcolor}         
\usepackage{graphicx}
\usepackage{subcaption}
\usepackage{amsmath}
\usepackage{enumitem}
\usepackage{array}

\title{MCP-Atlas: A Large-Scale Benchmark for Tool-Use Competency with Real MCP Servers}

\usepackage{verbatim}

\author[$\S$,*]{Chaithanya Bandi}
\author[$\dagger$,*]{Razvan-Gabriel Dumitru}
\author[$\dagger$]{Ben Hertzberg}
\author[$\dagger$]{Divyansh Agarwal}
\author[$\dagger$]{Geobio Boo}
\author[$\dagger$]{Tejas Polakam}
\author[$\dagger$]{Sami Hassaan}
\author[$\dagger$]{Jeff Da}
\author[$\dagger$]{HiJae Kim}
\author[$\dagger$]{Vipul Gupta}
\author[$\dagger$]{Manasi Sharma}
\author[$\dagger$]{Andrew Park}
\author[$\dagger$]{Martin Dimakis}
\author[$\dagger$]{Ernesto Gabriel Hernández Montoya}
\author[$\dagger$]{Dan Rambado}
\author[$\dagger$]{Ivan Salazar}
\author[$\dagger$]{Rafael Cruz}
\author[$\dagger$]{MohammadHossein Rezaei}
\author[$\dagger$]{Chetan Rane}
\author[$\dagger$]{Ben Levin}
\author[$\dagger$]{Daniel Yue Zhang}
\author[$\dagger$]{Brad Kenstler}
\author[$\dagger$]{Bing Liu}

\affil[$\dagger$]{Scale AI}
\affil[$\S$]{National University of Singapore}

\affil[ ]{*These authors contributed equally to this work}

\begin{document}

\maketitle

\begin{abstract}
The Model Context Protocol (MCP) is emerging as a standard interface through which large language model (LLM) agents discover and invoke external tools. 
However, existing MCP evaluations fall short along three key axes: realistic multi-step workflows with cross-server orchestration, breadth across authentic MCP servers rather than mocks, and structured, reproducible claim-level scoring disentangled from agent verbosity or style.
We introduce MCP-Atlas, a benchmark for measuring tool-use competency against production MCP servers. MCP-Atlas contains 1{,}000 natural-language tasks written and verified by human experts spanning 36 real MCP servers and 220 tools. Prompts do not specify servers, tools, or parameters, requiring agents to identify relevant tools among semantically plausible distractors and to compose multi-step, cross-server workflows. Each task is scored with a claim-level rubric, where final answers are scored against atomic factual claims grounded in tool outputs. This answer-centric scoring permits valid alternative tool-call trajectories to receive credit. We pair this with an 11-category diagnostic taxonomy that disentangles tool-call failures from cognitive failures in task understanding, synthesis, parsing, and stopping. Evaluating 20 frontier models from six providers under matched task-level conditions, we find pass rates up to 82.2\% at a 0.75 claim coverage threshold and a clear three-tier performance structure. Automated diagnostics show that 63.3\% of diagnosed failures are cognitive rather than tool-call related. Notably, several high-performing models fail after successful tool execution due to premature stopping or incorrect synthesis. We release the task schema, containerized harness, claim evaluator, and a 500-task public split, while reserving a 500-task private split to preserve leaderboard integrity. The code is at https://github.com/scaleapi/mcp-atlas.

\end{abstract}

\section{Introduction}

Large Language Models are increasingly deployed as agents that plan, invoke, and integrate external tools to accomplish complex user objectives. 
The Model Context Protocol (MCP)~\citep{anthropic2024mcp, mcp2025spec} has become the standard usage for this tool-use layer, providing uniform interfaces for server discovery, tool enumeration, typed parameter passing, and result consumption. Despite the rapid growth of MCP-based deployments, credibly evaluating such agents remains a significant challenge.

Practical deployments demand evaluations that simultaneously address three requirements: (i)~realistic, multi-step workflows that involve branching and cross-server orchestration, (ii)~breadth across authentic MCP servers and APIs rather than mocked substitutes, and (iii)~structured, reproducible, claim-level scoring that is disentangled from agent verbosity or stylistic preferences. Existing MCP-based benchmarks make progress along one or two of these but exhibit limitations on the third.

We introduce \textit{MCP-Atlas}, a large-scale real-server benchmark designed to measure tool-use competency under realistic single-turn prompts that require complex multi-tool workflows. Each of the 1,000 tasks executes against live MCP endpoints and defines a set of atomic factual claims that any correct answer should satisfy. This claim-level rubric decouples \emph{success} from adherence to a specific trajectory, so alternative tool-call paths that yield the same factual claims receive full credit, while also supporting post-hoc diagnosis through an 11-category taxonomy.

Beyond overall task success rate, MCP-Atlas exposes a controlled tool set per task (6--37 tools, of which only 2--8 are relevant), mixing required tools with plausible distractors from semantically similar categories. This design directly targets the ``unknown-tools'' challenge identified by prior work~\citep{luo2025mcpuniverse, liu2025mcpeval}, where agents must first discover which tools are relevant before invoking them. The benchmark spans 36 servers and 220 tools drawn from five application domains. The vast majority of tasks (98.6\%) require orchestration across two or more servers, and most tasks involve multi-hop dependencies where later tool calls are parameterized by earlier outputs.

\begin{figure}[t]
\centering
\includegraphics[width=\textwidth]{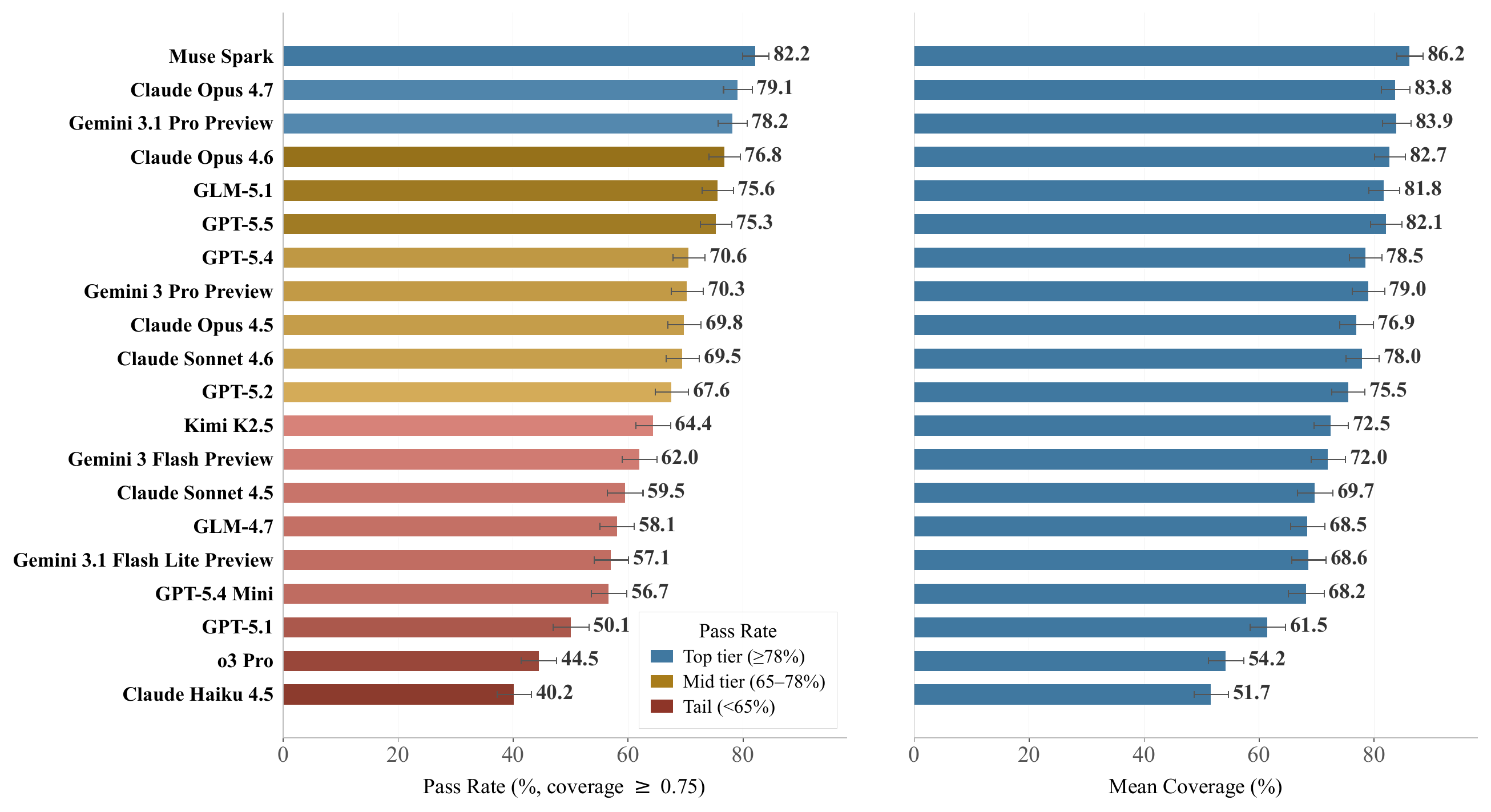}
\caption{Overall performance across 20 frontier models on the full 1{,}000-task set with 95\% confidence intervals. The best-performing model (Muse Spark) reaches 82.2\%, followed by Claude Opus 4.7 (79.1\%) and Gemini 3.1 Pro Preview (78.2\%). The open-source GLM-5.1 (75.6\%) is the highest-ranked open model in this evaluation. (exact reasoning config available in Appendix~\ref{app:model-reasoning-config}) } 
\label{fig:leaderboard_overall}
\end{figure}

Experiments on 20 frontier models from four proprietary and two open-source model providers reveal a clear three-tier structure (Figure~\ref{fig:leaderboard_overall}). A top cluster of three models sits between 78.2\% and 82.2\% pass rate within the confidence intervals, a mid-tier of eight models spans 67.6--76.8\%, and a long tail of nine models extends down to 40.2\%. Notably, the open-source GLM-5.1 reaches 75.6\%, entering the top empirical band in this evaluation that was previously exclusive to proprietary models. At the same time, o3 Pro, a leading reasoning model on mathematical and coding benchmarks, finishes near the bottom at 44.5\% because it produces no tool calls on 40\% of its failed tasks.

\textbf{Contributions.} The main contributions of this work are:
\begin{itemize}
    \item \textbf{A large-scale, real-server benchmark.} MCP-Atlas comprises 1{,}000 tasks across 36 production MCP servers and 220 tools. Tasks use natural-language prompts that avoid naming tools and reliably elicit multi-step behavior, with 98.6\% of tasks requiring cross-server orchestration.
    \item \textbf{A public release of data.} We open source a 500-task public subset that preserves the domain and complexity distribution of the full benchmark. The remaining 500 tasks form a held-out private split used for exposure monitoring and future leaderboard integrity.
    \item \textbf{Claims-based scoring with a fine-grained diagnostic taxonomy.} 
    Each failed task also receives an automated diagnosis from an 11-category taxonomy that distinguishes tool-call failures (malformed parameters, wrong tool selection, no tool use, failed error recovery) from cognitive failures (task misunderstanding, faulty synthesis, response misparsing, early termination, hallucinated facts, logical errors, constraint violations). Aggregate failure-mode statistics are released with the benchmark.
    \item \textbf{A large-scale empirical study.} We evaluate 20 frontier models, revealing a three-tier performance structure, a cognitive-over-tool failure split (63.3\% versus 36.7\%), and a striking tool-use policy mismatch in which strong reasoning models refuse to invoke tools on a substantial fraction of failed tasks.
\end{itemize}


\begin{table*}[t]
\centering
\small
\caption{Positioning MCP-Atlas among contemporary MCP evaluations. Scale metrics include number of servers (\#S), tools (\#T), and tasks (\#Tasks). For Distractors column: $\checkmark$ = systematic distractors included, blank = no distractors.}
\label{tab:mcp_comparison}
\resizebox{\textwidth}{!}{
\begin{tabular}{lccccccc}
\toprule
\textbf{Benchmark} &
\textbf{\#S} &
\textbf{\#T} &
\textbf{\#Tasks} &
\textbf{Cross-Server} &
\textbf{Distractors} &
\textbf{Natural Lang.} &
\textbf{Real Servers} \\
\midrule
MCP-Universe \citep{luo2025mcpuniverse} & 11 & 133 & 231 &  $\checkmark$ & $\checkmark$ & $\checkmark$ & $\checkmark$ \\
MCPEval \citep{liu2025mcpeval} & 5 & 19 & 676 &  Partial & & Partial & $\checkmark$ \\
MCP-Bench \citep{wang2025mcpbench} & 28 & 250 & 104  & $\checkmark$ & $\checkmark$ & $\checkmark$ & $\checkmark$ \\
Toolathlon \citep{li2025toolathlon} & 32 & 604 & 108  & $\checkmark$ & $\checkmark$ & $\checkmark$ & $\checkmark$ \\
MCPMark \citep{wu2025mcpmark} & 5 & $\sim$50 & 127  & & & $\checkmark$ & $\checkmark$ \\
LiveMCPBench \citep{mo2025livemcpbench} & 70 & 527 & 95  & $\checkmark$ & $\checkmark$ & $\checkmark$ & $\checkmark$ \\
MCPVerse \citep{zhao2025mcpverse} & 65 & 552 & --  & $\checkmark$ & $\checkmark$ & $\checkmark$ & Mixed \\
MCP-RADAR \citep{gao2025mcp} & 42 & -- & 507  & Partial & & Mixed & Mixed \\
\midrule
\textbf{MCP-Atlas (Ours)} & \textbf{36} & \textbf{220} & \textbf{1,000}  & $\checkmark$ & $\checkmark$ & $\checkmark$ & $\checkmark$ \\
\bottomrule
\end{tabular}
}
\end{table*}

\section{Related Work}
\label{sec:related_work}

\subsection{The Evolution from Static to Agentic Evaluation}

The evaluation landscape for LLMs has shifted significantly from static, knowledge-centric assessments to dynamic, interactive paradigms that measure agentic capabilities. Early benchmarks, such as MMLU~\citep{hendrycks2021mmlu} and HELM~\citep{liang2022helm}, focused on fixed-question sets evaluating factual recall and reasoning in isolated settings. While foundational, these benchmarks fail to capture the demands of real-world deployment, where models must interact with environments, manage uncertainty, and adapt plans dynamically.

Subsequent work introduced interactive evaluations emphasizing tool use and environmental interaction. Web-based benchmarks like WebArena~\citep{zhou2023webarena} and MiniWoB++~\citep{liu2018reinforcement} require agents to navigate and manipulate GUIs for tasks such as information retrieval or online shopping. OS-level suites, including OSWorld~\citep{xie2024osworld} and Android Arena~\citep{chai2025a3}, extend this to operating systems, testing multimodal control and long-horizon planning.

API- and function-calling benchmarks further advanced the field. ToolBench~\citep{qin2023toolllm} aggregates thousands of RESTful APIs for multi-step tasks. BFCL~\citep{patil2024bfcl} provides leaderboards focusing on function-calling with synthetic and real APIs. $\tau^2$-Bench~\citep{yao2024lambda} emphasizes retail and airline domains with interleaved user-agent interactions. Benchmarks like SWE-Bench~\citep{jimenez2023swebench} (software engineering) and GAIA~\citep{mialon2023gaia} (general assistance) incorporate diverse tools (code execution, web search), highlighting challenges in planning, error recovery, and orchestration.

\subsection{MCP-Based Evaluation}
\label{sec:mcp_eval}

The Model Context Protocol (MCP)~\citep{anthropic2024mcp, mcp2025spec} standardized tool declarations and server composition, spurring benchmarks that better reflect production environments. Table~\ref{tab:mcp_comparison} summarizes existing work. While these efforts have advanced the field, they reveal a persistent tension: benchmarks optimizing for evaluation rigor tend to remain small, while those achieving scale often compromise on objectivity or realism. MCP-Atlas is designed to resolve this tension.

Early MCP benchmarks prioritized rigorous, programmatic evaluation. MCP-Universe~\citep{luo2025mcpuniverse} introduced format, static, and dynamic evaluators with real-time ground truth verification, identifying the critical \textit{unknown-tools} challenge that motivates our distractor design. Toolathlon~\citep{li2025toolathlon} and MCPMark~\citep{wu2025mcpmark} similarly employ dedicated verification scripts against live software environments. However, the cost of manual task creation and custom verifiers limits these benchmarks to under 250 tasks---insufficient for statistically robust comparisons across the diversity of tool-use scenarios agents encounter in practice.

Efforts to scale have introduced different trade-offs. MCPEval~\citep{liu2025mcpeval} reaches 676 tasks through automated generation, but this can compromise task quality and naturalism. MCP-Bench~\citep{wang2025mcpbench} and LiveMCPBench~\citep{mo2025livemcpbench} adopt holistic LLM-as-judge scoring, which enables scale but introduces style biases---verbose responses may score differently than terse correct answers, reducing reproducibility. MCPVerse~\citep{zhao2025mcpverse} and MCP-RADAR~\citep{gao2025mcp} expand server coverage but rely partly on synthetic or mock implementations, limiting how well results predict real-world performance.

MCP-Atlas addresses these trade-offs through three design choices. First, we achieve scale (1,000 tasks across 36 servers and 220 tools) while maintaining quality through systematic manual verification rather than automated generation. Second, we replace holistic LLM-as-judge scoring with \textit{claims-based evaluation}: each task defines independent, verifiable factual claims that the correct answer must contain, enabling claim-level judge-assisted partial-credit scoring without the authoring burden of full programmatic validation. Third, we ensure evaluation fidelity by using exclusively real MCP servers and systematically include plausible distractors per task, directly testing the tool discovery capabilities that prior work identifies as a primary failure mode. Together, these choices yield a benchmark that is simultaneously large-scale, structured, and realistic.


\section{Benchmark Design}
\label{sec:benchmark-design}

MCP-Atlas is a 1{,}000-task evaluation suite built on 36 production Model Context Protocol (MCP) servers~\citep{mcp2025spec,anthropic2024mcp} exposing 220 tools across five application domains. Unlike benchmarks that rely on mocked APIs~\citep{zhao2025mcpverse,gao2025mcp} or holistic LLM-as-judge scoring~\citep{wang2025mcpbench,mo2025livemcpbench}, MCP-Atlas targets three threats to evaluation validity that prior work leaves uncontrolled: (i)~sanitized server behavior that suppresses real-world error modes, (ii)~trajectory-matching metrics that penalize valid alternative solutions, and (iii)~clean tool sets that obviate genuine tool discovery.
Tasks are randomly split 500/500 into public and private subsets for reproducibility and contamination monitoring, respectively.

\subsection{Server Ecosystem}
\label{sec:servers}

The 36 servers span five environment categories (Table~\ref{tab:env-buckets}). All are production MCP implementations, not synthetic stubs. Tasks execute against live endpoints that return authentic rate limits, pagination boundaries, schema-version mismatches, and transient error codes, capturing failure conditions that mocked environments elide by design. Servers are containerized with sandboxed filesystems and allow-listed network egress, version-pinned, and restarted between evaluation runs for reproducibility. The full server inventory and container configuration are in Appendix~\ref{app:servers}.

\begin{table}[t]
\centering
\caption{Environment buckets. Task share across five server categories with representative servers and typical pitfalls each category exercises.}
\label{tab:env-buckets}
\small
\setlength{\tabcolsep}{3.5pt}
\renewcommand{\arraystretch}{1.1}
\begin{tabular}{@{}llll@{}}
\toprule
\textbf{Bucket} & \textbf{Share} & \textbf{Representative Servers} & \textbf{Typical Pitfalls} \\
\midrule
Basic        & 30--35\% & brave\_search, exa, weather, maps & Query formulation, pagination \\
Productivity & 20--25\% & filesystem, notion, slack, arxiv   & Path scoping, rate limits \\
Coding       & 20--25\% & git, github, code-executor, cli    & Stateful ops, diff scope \\
Analytics    & 10--15\% & airtable, mongodb          & Schema alignment, typing \\
Financial    & 10--15\% & twelvedata, alchemy                 & Window alignment, symbols \\
\bottomrule
\end{tabular}
\end{table}

\subsection{Task Structure}
\label{sec:task-structure}

Each task exposes a tool set of 6--37 tools (mean 15.2), of which only 2--8 are required for the correct solution (mean 4.1). The remaining tools (mean 11.1 per task) are distractors drawn from semantically similar categories to test tool \emph{discovery} under noise, not just tool \emph{execution} with a known API. Tasks demand cross-server orchestration: 98.6\% require tools from two or more servers (mean 2.55), with 51.7\% needing exactly two, 38.2\% needing three, and 8.7\% needing four or more. Prompts are single-turn natural-language requests that avoid naming any server, tool, or parameter, forcing the model to infer the correct tool chain from task semantics alone.

Each task defines a \emph{claims list}: a set of atomic, independently verifiable factual statements grounded in tool outputs (mean 4.7 claims per task, range 1--23). An evaluator judge scores each claim as \emph{fulfilled}~(1.0), \emph{partially fulfilled}~(0.5), or \emph{not fulfilled}~(0.0). Task-level coverage is the mean claim score, and a task passes at coverage $\geq 0.75$ in the main analysis. A separate reference trajectory (mean 9.8 tool-call steps) records a minimal correct solution path for every task. 

In the released public split, this trajectory is serialized as an ordered sequence of tool calls, including tool names, arguments, dependencies, and task-specific returned evidence. The private split uses the same internal schema. We use reference trajectories for task solvability checks, tool-dependence validation, and failure diagnostics, but never for pass/fail scoring. Pass/fail is computed only from the model's final answer against the claim list, so alternative valid trajectories receive full credit. This decouples \emph{what} a model must achieve from \emph{how} it achieves it, so that valid alternative tool-call strategies receive full credit~\citep{patil2024bfcl}. Appendix~\ref{app:worked-example} walks through an end-to-end task drawn from the public split, showing the prompt, the enabled and required tools, the atomic claims list, and a passing model response under our rubric.

\subsection{Evaluation and Diagnostics}
\label{sec:eval-and-diagnostics}

All models are evaluated under matched task-level conditions: the same task prompt, exposed tool set, tool-call budget, scoring rubric, and final-answer protocol. Because provider APIs expose different native tool-calling interfaces, the low-level adapter is provider-specific, however, we apply no model-specific demonstrations, retry policies, task prompts, or scaffolding.

To decouple evaluator bias from model performance, three LLM judges independently score every task following the methodology of \citet{zheng2024judging}: Gemini~3.1~Pro~Preview~\citep{google2025gemini3} (primary), GPT-5.4~\citep{openai2025gpt5}, and Claude~Opus~4.6~\citep{anthropic2025claude4}. Model rankings have only small variations between them, with per-model score ranges of 2.1--4.6 percentage points (\S\ref{sec:judge-agreement}, Appendix~\ref{app:judge-comparison}). 

\label{sec:diagnostics}
Failed tasks receive an automated diagnosis assigning a primary failure mode from an 11-category taxonomy split into two families: \emph{tool-call issues} (malformed parameters, wrong tool selection, no tool use, failed error recovery) and \emph{cognitive issues} (task misunderstanding, faulty synthesis, response misparsing, early termination, hallucinated facts, logical errors, constraint violations). Table~\ref{tab:taxonomy-defs} in the Appendix clearly describes and provides an example for each failure mode.

\subsection{Quality Assurance}
\label{sec:qa}

Task and evaluation harness quality is maintained through 5 layers: (1)~expert execution of a reference trajectory against the configured MCP environment, (2)~domain-expert review for correctness and solvability, (3)~secondary review for tool and server name leakage in prompts, (4)~automated LLM verification of claims completeness and trajectory consistency, (5)~random-sample manual audit. The 500/500 public--private split enables exposure monitoring and leaderboard integrity checks. We report per-model performance gaps between splits in \S\ref{sec:contamination}. 

\section{Results and Analysis}
\label{sec:results}

We evaluate 20 models from six model families/providers: Anthropic, Google, OpenAI, Meta, Kimi~\citep{kimi2025k2}, and GLM~\citep{glm2024glm4}. Every model runs against all 1{,}000 tasks under identical conditions: the same exposed tools, tool-call budget, prompt template, and evaluator judge. Figure~\ref{fig:leaderboard_overall} reports pass rates with 95\% confidence intervals.

\subsection{Overall Performance}
\label{sec:overall-performance}

The leaderboard exhibits a clear three-tier structure. At the top, a cluster of three models sits within overlapping confidence intervals: Muse Spark (82.2\%), Claude Opus 4.7 (79.1\%), and Gemini 3.1 Pro Preview (78.2\%). Below this cluster, the mid-tier groups eight models spanning between 67\% and 77\% (Opus 4.6, GLM-5.1, GPT-5.5, GPT-5.4, Gemini 3 Pro Preview, Claude Opus 4.5, Claude Sonnet 4.6, GPT-5.2), forming a second tier. The remaining nine models fall below 65\%, with a long tail down to Claude Haiku 4.5 at 40.2\%.

Two results stand out. First, the open-source GLM-5.1 is the strongest open model in our study, reaching 75.6\% and entering the same empirical band as several top proprietary models under the primary judge, statistically tied with GPT-5.5 and above other GPT-5.x variants, which is within the confidence interval. This closes a gap that has long favored proprietary models on agentic tasks. 
Furthermore, as seen in our failure analysis (\S\ref{sec:failure-analysis}), 40.1\% of o3 Pro's failed trajectories contain no tool invocation, suggesting a mismatch between its default tool-use policy and tasks that require external evidence, despite being a leading reasoning model on mathematical and coding benchmarks.

\begin{table}[t]
\centering
\caption{Values show the percentage of each tool-call failure mode among all diagnosed failures for a given model. Tool\% is the share of all diagnosed failures that are tool-related.}
\label{tab:tool-call-failures}
\small
\setlength{\tabcolsep}{5pt}
\renewcommand{\arraystretch}{1.0}
\begin{tabular}{@{}lr rrrr@{}}
\toprule
\textbf{Model} & \textbf{Tool\%}
  & \textbf{Malformed}
  & \textbf{Wrong Tool}
  & \textbf{No Tool Use}
  & \textbf{Err. Recovery} \\
\midrule

\multicolumn{6}{@{}l}{\textit{Anthropic}} \\
\quad Claude Opus 4.7         & 34.6 & 11.8 &  1.7 &  9.1 & 12.0 \\
\quad Claude Opus 4.6         & 28.5 &  9.0 & 11.7 &  4.4 &  3.4 \\
\quad Claude Opus 4.5         & 29.9 &  6.4 &  8.8 &  4.6 & 10.2 \\
\quad Claude Sonnet 4.6       & 28.9 &  8.6 & 12.4 &  3.1 &  4.9 \\
\quad Claude Sonnet 4.5       & 37.3 &  6.4 &  8.8 &  6.4 & 15.8 \\
\quad Claude Haiku 4.5        & 46.7 &  5.7 &  9.6 & 25.7 &  5.7 
\\[4pt]
\multicolumn{6}{@{}l}{\textit{Google}} \\
\quad Gemini 3.1 Pro Preview  & 27.9 &  9.5 & 11.2 &  1.6 &  5.6 \\
\quad Gemini 3 Pro Preview    & 30.9 &  4.1 &  8.3 &  4.7 & 13.8 \\
\quad Gemini 3 Flash Preview  & 35.6 &  7.3 & 13.2 &  1.1 & 14.1 \\
\quad Gemini 3.1 Flash Lite   & 36.5 &  5.6 & 11.1 &  3.4 & 16.3 
\\[4pt]
\multicolumn{6}{@{}l}{\textit{OpenAI}} \\
\quad GPT-5.5                 & 33.2 & 11.6 &  4.0 & 13.0 &  4.6 \\
\quad GPT-5.4                 & 23.9 &  6.1 & 10.0 &  3.0 &  4.8 \\
\quad GPT-5.2                 & 33.6 &  4.5 &  7.5 &  8.8 & 12.9 \\
\quad GPT-5.4 Mini            & 29.5 &  2.7 &  7.4 &  6.2 & 13.2 \\
\quad GPT-5.1                 & 50.4 &  6.4 &  8.6 & 22.9 & 12.5 \\
\quad o3 Pro                  & 57.6 &  3.9 &  6.6 & 40.1 &  6.9 
\\[4pt]
\multicolumn{6}{@{}l}{\textit{Meta}} \\
\quad Muse Spark              & 27.1 &  5.2 &  8.9 &  1.8 & 11.1 
\\[4pt]
\multicolumn{6}{@{}l}{\textit{Open-Source}} \\
\quad GLM-5.1                 & 37.3 & 13.9 &  3.4 &  9.2 & 10.8 \\
\quad GLM-4.7                 & 45.9 &  8.0 & 11.2 &  7.9 & 18.8 \\
\quad Kimi K2.5               & 36.3 &  7.6 & 14.1 &  8.7 &  6.0 \\

\midrule
\textbf{Overall}              & \textbf{36.7} & \textbf{6.9} & \textbf{9.0} & \textbf{10.5} & \textbf{10.3} \\
\bottomrule
\end{tabular}

\vspace{2pt}
\footnotesize
\end{table}

\subsection{Failure Mode Analysis}
\label{sec:failure-analysis}



Across the 20 models, 63.3\% of roughly 6,900 diagnosed failures are cognitive and 36.7\% are tool-related (Tables~\ref{tab:tool-call-failures}, \ref{tab:cognitive-failures}). This aggregate split reveals that MCP-Atlas is not merely testing whether models can issue valid tool calls. The dominant failure for current agents is often what happens after tool access: understanding the task, deciding whether enough evidence has been gathered, and synthesizing the final answer from tool outputs. Tool-call failures still matter, especially in the tail of the leaderboard, but they no longer explain most remaining errors.

The clearest tool-use breakdown occurs in lower-performing models. o3 Pro is the most striking case: 57.6\% of its diagnosed failures are tool-related, and 40.1\% of its failed trajectories contain no tool invocation at all, despite tasks requiring external evidence. GPT-5.1 and Claude Haiku 4.5 show similar, though less extreme, acquisition failures, with tool-related shares of 50.4\% and 46.7\%, respectively. These models often fail before entering the gathering evidence, suggesting that their default tool-use policies are poorly calibrated for MCP with unknown tools and realistic distractors.

Among stronger models, the bottleneck shifts from calling tools to completing and using the evidence chain, and the newest frontier models appear to move this boundary further downstream. Early termination is the signature failure for one high-performing subgroup: Gemini 3.1 Pro Preview, GPT-5.4, and Claude Opus 4.6 terminate early on 42.8\%, 41.4\%, and 36.3\% of their diagnosed failures, respectively. In contrast, the newer GPT-5.5 and Claude Opus 4.7 almost eliminate this mode, with early-termination rates of only 3.6\% and 3.8\%, but their errors reappear as harder post-retrieval failures: faulty synthesis rises to 19.8\% for GPT-5.5 and 25.0\% for Claude Opus 4.7. This suggests an emerging frontier regime in which agents increasingly persist through the workflow, but still miscombine, misinterpret, or misreport the evidence they collect. The next gains on MCP-Atlas therefore likely require not only claim-aware stopping criteria, but also trajectory-grounded verification of the final answer against the collected tool evidence.

\begin{table}[t]
\centering
\caption{Values show the percentage of each cognitive failure mode among all diagnosed failures for a given model. Cog\% is the share of all diagnosed failures that are cognitive.}
\label{tab:cognitive-failures}
\small
\setlength{\tabcolsep}{3.2pt}
\renewcommand{\arraystretch}{1.0}
\begin{tabular}{@{}lr rrrrrrr@{}}
\toprule
\textbf{Model} & \textbf{Cog\%}
  & \rotatebox[origin=bl]{60}{\textbf{Task Misund.}}
  & \rotatebox[origin=bl]{60}{\textbf{Faulty Synth.}}
  & \rotatebox[origin=bl]{60}{\textbf{Misparsing}}
  & \rotatebox[origin=bl]{60}{\textbf{Early Term.}}
  & \rotatebox[origin=bl]{60}{\textbf{Hallucinated}}
  & \rotatebox[origin=bl]{60}{\textbf{Logical Err.}}
  & \rotatebox[origin=bl]{60}{\textbf{Constraint}} \\
\midrule

\multicolumn{9}{@{}l}{\textit{Anthropic}} \\
\quad Claude Opus 4.7         & 65.4 & 13.9 & 25.0 &  4.8 &  3.8 & 7.9 & 8.9 & 1.0 \\
\quad Claude Opus 4.6         & 71.5 & 14.0 &  8.7 &  4.8 & 36.3 & 2.1 & 3.0 & 2.5 \\
\quad Claude Opus 4.5         & 70.1 & 16.2 & 15.6 &  9.0 & 15.8 & 6.0 & 5.2 & 2.4 \\
\quad Claude Sonnet 4.6       & 71.1 & 16.9 & 13.6 & 14.9 & 11.0 & 6.9 & 5.1 & 2.6 \\
\quad Claude Sonnet 4.5       & 62.7 & 14.3 & 12.0 & 10.1 & 17.7 & 3.6 & 3.4 & 1.5 \\
\quad Claude Haiku 4.5        & 53.3 & 10.2 &  8.0 &  7.7 & 17.0 & 4.2 & 4.1 & 2.1 \\[4pt]

\multicolumn{9}{@{}l}{\textit{Google}} \\
\quad Gemini 3.1 Pro Preview  & 72.1 & 13.7 &  7.2 &  3.0 & 42.8 & 1.6 & 2.6 & 1.2 \\
\quad Gemini 3 Pro Preview    & 69.1 & 17.3 & 12.8 &  8.9 & 18.6 & 5.6 & 3.5 & 2.3 \\
\quad Gemini 3 Flash Preview  & 64.4 & 14.2 & 10.5 & 10.7 & 14.1 & 8.5 & 4.3 & 2.1 \\
\quad Gemini 3.1 Flash Lite   & 63.5 & 13.1 &  9.9 & 10.5 & 20.8 & 3.4 & 3.1 & 2.7 \\[4pt]

\multicolumn{9}{@{}l}{\textit{OpenAI}} \\
\quad GPT-5.5                 & 66.8 & 23.4 & 19.8 &  6.2 &  3.6 & 1.0 &11.2 & 1.6 \\
\quad GPT-5.4                 & 76.1 & 15.0 & 11.2 &  4.8 & 41.4 & 0.5 & 2.0 & 1.1 \\
\quad GPT-5.2                 & 66.4 & 19.1 & 12.3 &  5.8 & 22.6 & 3.0 & 2.2 & 1.3 \\
\quad GPT-5.4 Mini            & 70.5 & 18.5 & 13.0 &  7.0 & 27.4 & 2.8 & 1.2 & 0.6 \\
\quad GPT-5.1                 & 49.6 & 12.0 &  7.9 &  7.1 & 16.5 & 2.6 & 2.0 & 1.5 \\
\quad o3 Pro                  & 42.4 &  9.0 &  4.5 &  4.6 & 17.2 & 4.6 & 1.8 & 0.6 \\[4pt]

\multicolumn{9}{@{}l}{\textit{Meta}} \\
\quad Muse Spark              & 72.9 & 17.2 & 16.9 & 11.7 & 11.4 & 7.1 & 5.2 & 3.4 \\[4pt]

\multicolumn{9}{@{}l}{\textit{Open-Source}} \\
\quad GLM-5.1                 & 62.7 & 15.5 & 22.5 &  4.0 &  3.8 & 6.7 & 8.8 & 1.3 \\
\quad GLM-4.7                 & 54.1 & 16.8 &  8.2 &  7.0 & 14.1 & 2.2 & 4.2 & 1.7 \\
\quad Kimi K2.5               & 63.7 & 17.1 & 12.1 &  9.9 & 13.7 & 4.7 & 4.7 & 1.4 \\

\midrule
\textbf{Overall}              & \textbf{63.3} & \textbf{15.1} & \textbf{12.0} & \textbf{7.6} & \textbf{18.7} & \textbf{4.1} & \textbf{4.1} & \textbf{1.7} \\
\bottomrule
\end{tabular}

\vspace{2pt}
\footnotesize
\end{table}

\subsection{Efficiency Frontier}
\label{sec:efficiency}

\begin{figure}[t]
\centering
\includegraphics[width=\textwidth]{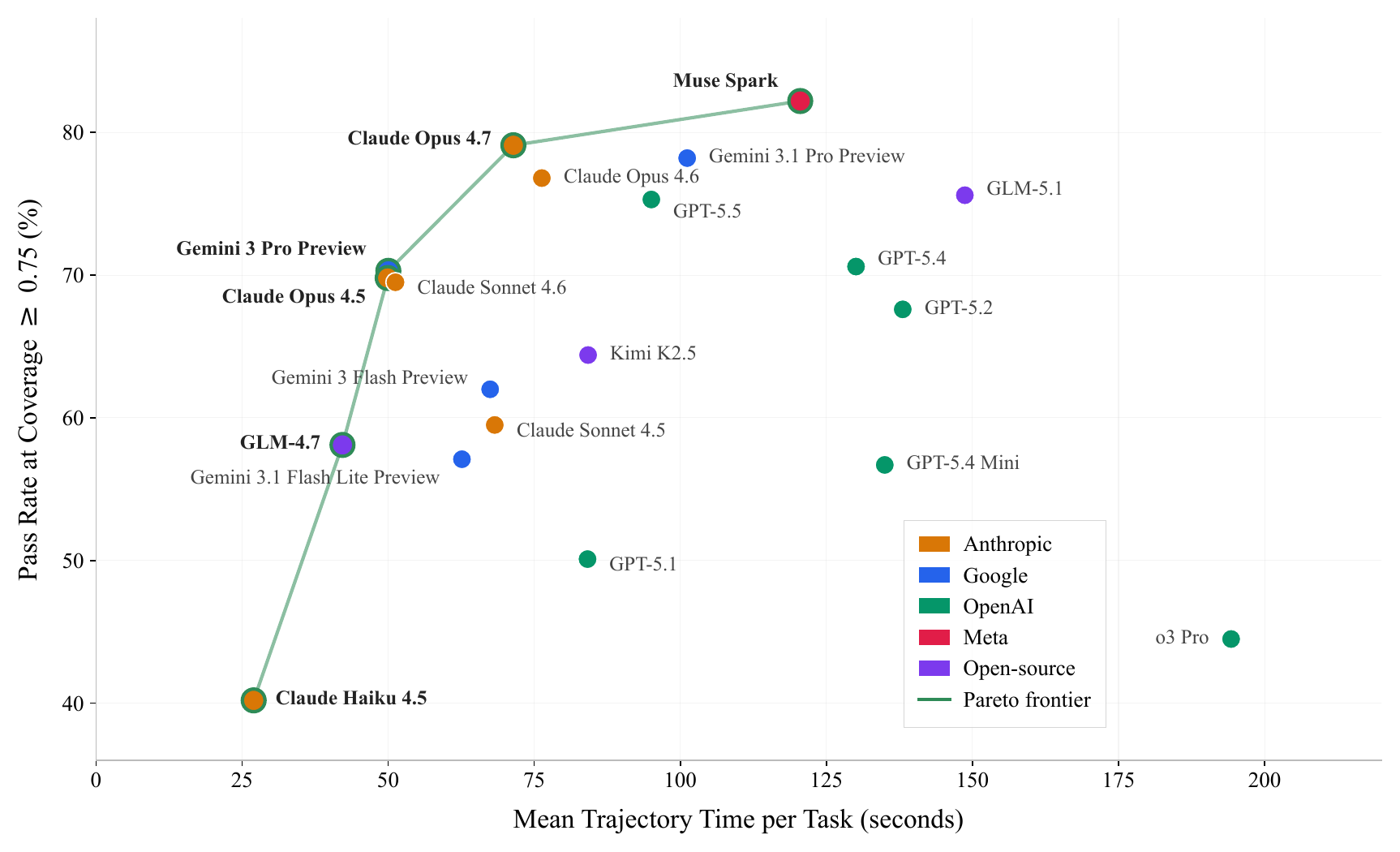}
\caption{Efficiency Pareto frontier. Each point is a model, positioned by mean wall-clock trajectory time per task (x-axis) and pass rate (y-axis). Frontier models are bolded and connected by the green line. OpenAI's reasoning-heavy configurations are strictly dominated by faster alternatives.}
\label{fig:efficiency_pareto}
\end{figure}

Beyond raw accuracy, practitioners care about how much time a given pass rate takes. Figure~\ref{fig:efficiency_pareto} plots mean wall-clock trajectory time per task against pass rate for every model. Six models sit on the Pareto frontier: Claude Haiku 4.5 at the fast-but-inaccurate end (27s, 40.2\%), followed by GLM-4.7 (42s, 58.1\%), Claude Opus 4.5 and Gemini 3 Pro Preview (both near 50s, 70\%), Claude Opus 4.7 (71s, 79.1\%), and Muse Spark at the top (121s, 82.2\%).

Several models are \emph{strictly dominated} by faster alternatives. o3 Pro, at 194s and 44.5\%, is the slowest model in the set yet also among the least accurate, trailing every other model in the lower half of the leaderboard by a wide margin in wall-clock efficiency. GPT-5.1 (84s, 50.1\%) is dominated by GLM-4.7, which is both faster and more accurate. GPT-5.4 Mini (135s, 56.7\%) is slower than its larger counterpart GPT-5.4 (130s, 70.6\%) while also scoring lower, indicating that the ``Mini'' designation does not translate into faster execution in agentic settings where wall-clock time is dominated by tool-call roundtrip latency rather than model inference. The open-source GLM-5.1 is not on the frontier despite its strong pass rate, because it averages 149s per task, roughly twice the wall-clock cost of Claude Opus 4.6 and Opus 4.7 for comparable accuracy. 

\subsection{Cross-Judge Agreement}
\label{sec:judge-agreement}

Three LLM judges score all 1{,}000 tasks for every model: Gemini 3.1 Pro Preview (primary), GPT-5.4, and Claude Opus 4.6. The results are presented in Appendix~\ref{app:judge-comparison}. The primary judge is the most generous (mean pass rate 65.4\%), GPT-5.4 the strictest (62.3\%), and Claude Opus 4.6 intermediate (63.9\%). Per-model score ranges span 2.1 to 4.6 percentage points. 
Model rankings are broadly preserved across judges, but the few changes that occur are concentrated in the crowded upper cluster. The largest movement is Claude Opus 4.7, which ranks second under Gemini 3.1 Pro Preview, third under GPT-5.4, and fifth under Claude Opus 4.6. We do not interpret this as self-preference: the Claude judge does not favor the newer Anthropic model and instead places Gemini 3.1 Pro Preview and GLM-5.1 above it. A more conservative interpretation is that several models in the upper cluster are separated by margins comparable to the 2.1--4.6 percentage-point cross-judge range. Thus, small rank swaps near the top should be read as evaluator uncertainty rather than robust capability differences. This supports using a single primary judge for the main leaderboard while reporting cross-judge ranges as calibration.


\subsection{Public--Private Split Sensitivity}
\label{sec:contamination}

The 500/500 public--private split supports reproducibility while preserving a held-out slice for leaderboard integrity. Because the released public split includes prompts, tool sets, claims, and reference trajectories, gaps after release are best interpreted as split-sensitivity and exposure-monitoring signals, not as direct evidence of memorization. Most models score higher on public tasks (Figure 3), with gaps from $-0.8$ to $+9.6$ pp, and the gap does not track overall pass rate. The gaps show visible model-family/provider clustering, but because this is a small model-level comparison with uneven group sizes, we treat the pattern as an audit signal rather than evidence of contamination. OpenAI models cluster at +8.4 pp (range +7.0 to +9.6), Google averages +5.2 pp, Anthropic +4.8 pp, and open-source models +3.8 pp, while Kimi K2.5 is the only model with higher private than public performance ($-0.8$ pp). We therefore report split gaps as audit statistics alongside overall score.


\begin{figure}[t]
\centering
\includegraphics[width=\textwidth]{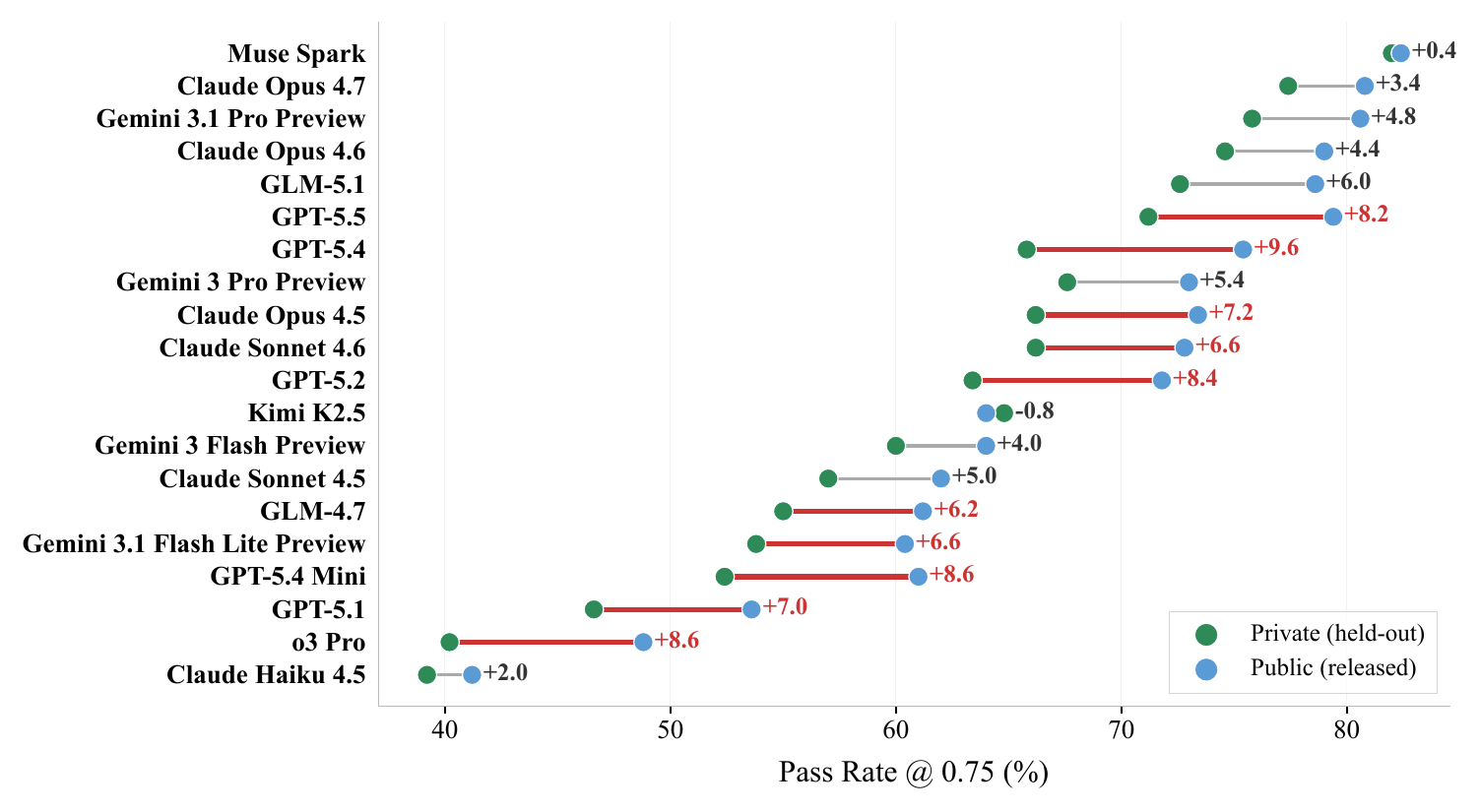}
\caption{Public versus private split performance for all 20 models, ordered by overall pass rate. Red connectors indicate gaps greater than 6 percentage points. These gaps are audit triggers for possible benchmark exposure, split sensitivity, or provider-specific evaluation effects. They are not direct evidence of contamination. Kimi K2.5 is the only model where private performance exceeds public.}
\label{fig:public_private_split}
\end{figure}

\section{Discussion}
\label{sec:discussion}





MCP-Atlas shows that real-server MCP evaluation is not saturated, but it is no longer explained by function-calling mechanics alone. The frontier is compressed: the top three models span 78.2--82.2\% pass rate, and cross-judge ranges are only 2.1--4.6 points, so small swaps near the top should be interpreted as evaluator uncertainty rather than stable capability gaps. At the same time, the best model still fails nearly one in five tasks, leaving headroom on realistic multi-server workflows.

The diagnostic results identify where that headroom lies. Tool-call failures dominate some tail models, especially systems that often never invoke tools, but across all diagnosed failures nearly two thirds are cognitive. Stronger agents usually enter the correct evidence-gathering regime, then stop right 
before covering all claims or synthesize the collected outputs incorrectly. This shifts the research target from valid API calls alone toward claim-aware stopping rules and trajectory-grounded answer verification.


Finally, MCP-Atlas exposes deployment-relevant tradeoffs and limitations hidden by a single pass-rate: several models are accuracy--latency dominated, split gaps vary by provider, and results reflect live-server drift, judge uncertainty, and provider-specific harness defaults. Future work should report a diagnostic profile---accuracy, efficiency, judge sensitivity, split gap, drift, and failure mode---and add multi-turn clarification and distractor ablations.


\section{Conclusion}
\label{sec:conclusion}

We presented MCP-Atlas, a 1{,}000-task benchmark that evaluates tool-use competency against 36 production MCP servers and 220 real tools, scored by a partial-credit claims-based rubric and evaluated across three independent LLM judges. Across 20 frontier models, we find that the agentic frontier is no longer bottlenecked by tool mechanics. Tool-related failures shrink predictably as capability rises, and among top models nearly two of every three remaining failures are cognitive, with a striking concentration in early termination, faulty synthesis, and no-tool-use failures in reasoning-trained checkpoints. We provide the 500-task public split, containerized harness, claims evaluator, scoring pipeline, and Croissant/RAI metadata at submission. Our hope is that MCP-Atlas serves both as a leaderboard with enough headroom to track the next generation of agents and as a diagnostic instrument that points training and post-training research at the stages where current models break. 

\bibliographystyle{unsrtnat} 
\bibliography{iclr2026_conference}

@inproceedings{hendrycks2021mmlu,
  title = {Measuring Massive Multitask Language Understanding},
  author = {Dan Hendrycks and Collin Burns and Steven Basart and Andy Zou and Mantas Mazeika and Dawn Song and Jacob Steinhardt},
  booktitle = {International Conference on Learning Representations},
  year = {2021},
  url = {https://openreview.net/forum?id=d7KBjmI3GmQ}
}

@article{efron1979bootstrap,
    author  = {Efron, Bradley},
    title   = {Bootstrap Methods: Another Look at the Jackknife},
    journal = {The Annals of Statistics},
    volume  = {7},
    number  = {1},
    pages   = {1--26},
    year    = {1979}
  }

@article{liang2022helm,
  title = {Holistic Evaluation of Language Models},
  author = {Percy Liang and Rishi Bommasani and Tony Lee and Dimitris Tsipras and Dilara Soylu and Michihiro Yasunaga and Yian Zhang and Deepak Narayanan and Yuhuai Wu and Ananya Kumar and Benjamin Newman and Binhang Yuan and Bobby Yan and Ce Zhang and Christian Cosgrove and Christopher D. Manning and Christopher R{\'e} and Diana Acosta-Navas and Drew A. Hudson and Eric Zelikman and Esin Durmus and Faisal Ladhak and Frieda Rong and Hongyu Ren and Huaxiu Yao and Jue Wang and Keshav Santhanam and Laurel Orr and Lucia Zheng and Mert Yuksekgonul and Mirac Suzgun and Nathan Kim and Neel Guha and Niladri Chatterji and Omar Khattab and Peter Henderson and Qian Huang and Ryan Chi and Sang Michael Xie and Shibani Santurkar and Surya Ganguli and Tatsunori Hashimoto and Thomas Icard and Tianyi Zhang and Vishrav Chaudhary and William Wang and Xuechen Li and Yifan Mai and Yuhui Zhang and Yuta Koreeda},
  journal = {Transactions on Machine Learning Research},
  year = {2023},
  url = {https://openreview.net/forum?id=iO4LZibEqW},
}

@inproceedings{zhou2023webarena,
  title = {{WebArena}: A Realistic Web Environment for Building Autonomous Agents},
  author = {Shuyan Zhou and Frank F. Xu and Hao Zhu and Xuhui Zhou and Robert Lo and Abishek Sridhar and Xianyi Cheng and Tianyue Ou and Yonatan Bisk and Daniel Fried and Uri Alon and Graham Neubig},
  booktitle = {International Conference on Learning Representations},
  year = {2024},
  url = {https://openreview.net/forum?id=oKn9c6ytLx}
}

@inproceedings{liu2018reinforcement,
  title = {Reinforcement Learning on Web Interfaces using Workflow-Guided Exploration},
  author = {Evan Zheran Liu and Kelvin Guu and Panupong Pasupat and Tianlin Shi and Percy Liang},
  booktitle = {International Conference on Learning Representations},
  year = {2018},
  url = {https://openreview.net/forum?id=ryTp3f-0-}
}

@inproceedings{xie2024osworld,
  title = {{OSWorld}: Benchmarking Multimodal Agents for Open-Ended Tasks in Real Computer Environments},
  author = {Tianbao Xie and Danyang Zhang and Jixuan Chen and Xiaochuan Li and Siheng Zhao and Ruisheng Cao and Toh Jing Hua and Zhoujun Cheng and Dongchan Shin and Fangyu Lei and Yitao Liu and Yiheng Xu and Shuyan Zhou and Silvio Savarese and Caiming Xiong and Victor Zhong and Tao Yu},
  booktitle = {Advances in Neural Information Processing Systems, Datasets and Benchmarks Track},
  year = {2024},
  url = {https://openreview.net/forum?id=tN61DTr4Ed}
}

@article{chai2025a3,
  title = {{A3}: Android Agent Arena for Mobile GUI Agents},
  author = {Yuxiang Chai and Hanhao Li and Jiayu Zhang and Liang Liu and Guozhi Wang and Shuai Ren and Siyuan Huang and Hongsheng Li},
  journal = {arXiv preprint arXiv:2501.01149},
  year = {2025},
  url = {https://arxiv.org/abs/2501.01149}
}

@inproceedings{qin2023toolllm,
  title = {{ToolLLM}: Facilitating Large Language Models to Master 16000+ Real-World APIs},
  author = {Yujia Qin and Shihao Liang and Yining Ye and Kunlun Zhu and Lan Yan and Yaxi Lu and Yankai Lin and Xin Cong and Xiangru Tang and Bill Qian and Sihan Zhao and Lauren Hong and Runchu Tian and Ruobing Xie and Jie Zhou and Mark Gerstein and Dahai Li and Zhiyuan Liu and Maosong Sun},
  booktitle = {International Conference on Learning Representations},
  year = {2024},
  url = {https://openreview.net/forum?id=dHng2O0Jjr}
}

@inproceedings{patil2024bfcl,
  title = {The Berkeley Function Calling Leaderboard ({BFCL}): From Tool Use to Agentic Evaluation of Large Language Models},
  author = {Shishir G. Patil and Huanzhi Mao and Fanjia Yan and Charlie Cheng-Jie Ji and Vishnu Suresh and Ion Stoica and Joseph E. Gonzalez},
  booktitle = {Proceedings of the 42nd International Conference on Machine Learning},
  series = {Proceedings of Machine Learning Research},
  volume = {267},
  pages = {48371--48392},
  publisher = {PMLR},
  year = {2025},
  url = {https://proceedings.mlr.press/v267/patil25a.html}
}

@misc{yao2024lambda,
      title={$\tau^2$-Bench: Evaluating Conversational Agents in a Dual-Control Environment}, 
      author={Victor Barres and Honghua Dong and Soham Ray and Xujie Si and Karthik Narasimhan},
      year={2025},
      eprint={2506.07982},
      archivePrefix={arXiv},
      primaryClass={cs.AI},
      url={https://arxiv.org/abs/2506.07982}, 
}

@inproceedings{jimenez2023swebench,
  title = {{SWE}-bench: Can Language Models Resolve Real-world {GitHub} Issues?},
  author = {Carlos E. Jimenez and John Yang and Alexander Wettig and Shunyu Yao and Kexin Pei and Ofir Press and Karthik R. Narasimhan},
  booktitle = {International Conference on Learning Representations},
  year = {2024},
  url = {https://openreview.net/forum?id=VTF8yNQM66}
}

@inproceedings{mialon2023gaia,
  title = {{GAIA}: A Benchmark for General {AI} Assistants},
  author = {Gr{\'e}goire Mialon and Cl{\'e}mentine Fourrier and Craig Swift and Thomas Wolf and Yann LeCun and Thomas Scialom},
  booktitle = {International Conference on Learning Representations},
  year = {2024},
  url = {https://openreview.net/forum?id=fibxvahvs3}
}

@misc{mcp2025spec,
  title = {{Model Context Protocol} Specification},
  author = {{Model Context Protocol}},
  year = {2025},
  url = {https://modelcontextprotocol.io/specification/2025-03-26}
}

@misc{anthropic2024mcp,
  title = {Introducing the {Model Context Protocol}},
  author = {{Anthropic}},
  year = {2024},
  month = nov,
  url = {https://www.anthropic.com/news/model-context-protocol}
}

@article{luo2025mcpuniverse,
  title = {{MCP-Universe}: Benchmarking Large Language Models with Real-World {Model Context Protocol} Servers},
  author = {Ziyang Luo and Zhiqi Shen and Wenzhuo Yang and Zirui Zhao and Prathyusha Jwalapuram and Amrita Saha and Doyen Sahoo and Silvio Savarese and Caiming Xiong and Junnan Li},
  journal = {arXiv preprint arXiv:2508.14704},
  year = {2025},
  url = {https://arxiv.org/abs/2508.14704},
}

@inproceedings{wang2025mcpbench,
  title = {{MCP}-Bench: Benchmarking Tool-Using {LLM} Agents with Complex Real-World Tasks via {MCP} Servers},
  author = {Zhenting Wang and Qi Chang and Hemani Patel and Shashank Biju and Cheng-En Wu and Quan Liu and Aolin Ding and Alireza Rezazadeh and Ankit Shah and Yujia Bao and Eugene Siow},
  booktitle = {International Conference on Learning Representations},
  year = {2026},
  url = {https://openreview.net/forum?id=fe8mzHwMxN},
}

@inproceedings{liu2025mcpeval,
  title = {{MCPEval}: Automatic {MCP}-based Deep Evaluation for {AI} Agent Models},
  author = {Zhiwei Liu and Jielin Qiu and Shiyu Wang and Jianguo Zhang and Zuxin Liu and Roshan Ram and Haolin Chen and Weiran Yao and Shelby Heinecke and Silvio Savarese and Huan Wang and Caiming Xiong},
  booktitle = {Proceedings of the 2025 Conference on Empirical Methods in Natural Language Processing: System Demonstrations},
  pages = {373--402},
  address = {Suzhou, China},
  publisher = {Association for Computational Linguistics},
  doi = {10.18653/v1/2025.emnlp-demos.27},
  year = {2025},
  url = {https://aclanthology.org/2025.emnlp-demos.27/}
}

@article{mo2025livemcpbench,
  title = {{LiveMCPBench}: Can Agents Navigate an Ocean of {MCP} Tools?},
  author = {Guozhao Mo and Wenliang Zhong and Jiawei Chen and Qianhao Yuan and Xuanang Chen and Yaojie Lu and Hongyu Lin and Ben He and Xianpei Han and Le Sun},
  journal = {arXiv preprint arXiv:2508.01780},
  year = {2025},
  url = {https://arxiv.org/abs/2508.01780},
}

@article{gao2025mcp,
  title = {{MCP-RADAR}: A Multi-Dimensional Benchmark for Evaluating Tool Use Capabilities in Large Language Models},
  author = {Xuanqi Gao and Siyi Xie and Juan Zhai and Shqing Ma and Chao Shen},
  journal = {arXiv preprint arXiv:2505.16700},
  year = {2025},
  url = {https://arxiv.org/abs/2505.16700}
}

@inproceedings{li2025toolathlon,
  title = {The Tool Decathlon: Benchmarking Language Agents for Diverse, Realistic, and Long-Horizon Task Execution},
  author = {Junlong Li and Wenshuo Zhao and Jian Zhao and Weihao Zeng and Haoze Wu and Xiaochen Wang and Rui Ge and Yuxuan Cao and Yuzhen Huang and Wei Liu and Junteng Liu and Zhaochen Su and Yiyang Guo and Fan Zhou and Lueyang Zhang and Juan Michelini and Xingyao Wang and Xiang Yue and Shuyan Zhou and Graham Neubig and Junxian He},
  booktitle = {International Conference on Learning Representations},
  year = {2026},
  url = {https://openreview.net/forum?id=z53s5p0qhf},
}

@inproceedings{wu2025mcpmark,
  title = {{MCPMark}: A Benchmark for Stress-Testing Realistic and Comprehensive {MCP} Use},
  author = {Zijian Wu and Xiangyan Liu and Xinyuan Zhang and Lingjun Chen and Fanqing Meng and Lingxiao Du and Yiran Zhao and Fanshi Zhang and Yaoqi Ye and Jiawei Wang and Zirui Wang and Jinjie Ni and Yufan Yang and Arvin Xu and Michael Qizhe Shieh},
  booktitle = {International Conference on Learning Representations},
  year = {2026},
  url = {https://openreview.net/forum?id=uobROwBsJm},
}

@article{zhao2025mcpverse,
  title = {{MCPVerse}: An Expansive, Real-World Benchmark for Agentic Tool Use},
  author = {Fei Lei and Yibo Yang and Wenxiu Sun and Dahua Lin},
  journal = {arXiv preprint arXiv:2508.16260},
  year = {2025},
  url = {https://arxiv.org/abs/2508.16260},
}

@inproceedings{zheng2024judging,
  title = {Judging {LLM}-as-a-Judge with {MT}-Bench and Chatbot Arena},
  author = {Lianmin Zheng and Wei-Lin Chiang and Ying Sheng and Siyuan Zhuang and Zhanghao Wu and Yonghao Zhuang and Zi Lin and Zhuohan Li and Dacheng Li and Eric P. Xing and Hao Zhang and Joseph E. Gonzalez and Ion Stoica},
  booktitle = {Advances in Neural Information Processing Systems},
  year = {2024},
  url = {https://openreview.net/forum?id=uccHPGDlao}
}

@article{anthropic2025claude4,
  title = {The {Claude} Model Family},
  author = {{Anthropic}},
  year = {2025},
  url = {https://www.anthropic.com/research/claude-model-card}
}

@article{openai2025gpt5,
  title = {{GPT}-5 System Card},
  author = {{OpenAI}},
  year = {2025},
  url = {https://openai.com/index/gpt-5-system-card/}
}

@article{google2025gemini3,
  title = {Gemini 3 Technical Report},
  author = {{Google DeepMind}},
  year = {2025},
  url = {https://deepmind.google/technologies/gemini/}
}

@article{glm2024glm4,
  title = {{ChatGLM}: A Family of Large Language Models from {GLM}-130B to {GLM}-4 All Tools},
  author = {Team GLM and Aohan Zeng and Bin Xu and Bowen Wang and Chenhui Zhang and Da Yin and Diego Rojas and Guanyu Feng and Hanlin Zhao and Hanyu Lai and Hao Yu and Hongning Wang and Jiadai Sun and Jiajie Zhang and Jiale Cheng and Jiayi Gui and Jie Tang and Jing Zhang and Juanzi Li and Lei Zhao and Lindong Wu and Lucen Zhong and Mingdao Liu and Minlie Huang and Peng Zhang and Qinkai Zheng and Rui Lu and Shuaiqi Duan and Shudan Zhang and Shulin Cao and Shuxun Yang and Weng Lam Tam and Wenyi Zhao and Xiao Liu and Xiao Xia and Xiaohan Zhang and Xiaotao Gu and Xin Lv and Xinghan Liu and Xinyi Liu and Xinyue Yang and Xixuan Song and Xunhai Zhang and Yifan An and Yifan Xu and Yilin Niu and Yuantao Yang and Yueyan Li and Yushi Bai and Yuxiao Dong and Zehan Qi and Zhaoyu Wang and Zhen Yang and Zhengxiao Du and Zhenyu Hou and Zihan Wang},
  journal = {arXiv preprint arXiv:2406.12793},
  year = {2024},
  url = {https://arxiv.org/abs/2406.12793}
}

@misc{kimi2025k2,
      title={Kimi K2: Open Agentic Intelligence}, 
      author={Kimi Team and Yifan Bai and Yiping Bao and Y. Charles and Cheng Chen and Guanduo Chen and Haiting Chen and Huarong Chen and Jiahao Chen and Ningxin Chen and Ruijue Chen and Yanru Chen and Yuankun Chen and Yutian Chen and Zhuofu Chen and Jialei Cui and Hao Ding and Mengnan Dong and Angang Du and Chenzhuang Du and Dikang Du and Yulun Du and Yu Fan and Yichen Feng and Kelin Fu and Bofei Gao and Chenxiao Gao and Hongcheng Gao and Peizhong Gao and Tong Gao and Yuyao Ge and Shangyi Geng and Qizheng Gu and Xinran Gu and Longyu Guan and Haiqing Guo and Jianhang Guo and Xiaoru Hao and Tianhong He and Weiran He and Wenyang He and Yunjia He and Chao Hong and Hao Hu and Yangyang Hu and Zhenxing Hu and Weixiao Huang and Zhiqi Huang and Zihao Huang and Tao Jiang and Zhejun Jiang and Xinyi Jin and Yongsheng Kang and Guokun Lai and Cheng Li and Fang Li and Haoyang Li and Ming Li and Wentao Li and Yang Li and Yanhao Li and Yiwei Li and Zhaowei Li and Zheming Li and Hongzhan Lin and Xiaohan Lin and Zongyu Lin and Chengyin Liu and Chenyu Liu and Hongzhang Liu and Jingyuan Liu and Junqi Liu and Liang Liu and Shaowei Liu and T. Y. Liu and Tianwei Liu and Weizhou Liu and Yangyang Liu and Yibo Liu and Yiping Liu and Yue Liu and Zhengying Liu and Enzhe Lu and Haoyu Lu and Lijun Lu and Yashuo Luo and Shengling Ma and Xinyu Ma and Yingwei Ma and Shaoguang Mao and Jie Mei and Xin Men and Yibo Miao and Siyuan Pan and Yebo Peng and Ruoyu Qin and Zeyu Qin and Bowen Qu and Zeyu Shang and Lidong Shi and Shengyuan Shi and Feifan Song and Jianlin Su and Zhengyuan Su and Lin Sui and Xinjie Sun and Flood Sung and Yunpeng Tai and Heyi Tang and Jiawen Tao and Qifeng Teng and Chaoran Tian and Chensi Wang and Dinglu Wang and Feng Wang and Hailong Wang and Haiming Wang and Jianzhou Wang and Jiaxing Wang and Jinhong Wang and Shengjie Wang and Shuyi Wang and Si Wang and Xinyuan Wang and Yao Wang and Yejie Wang and Yiqin Wang and Yuxin Wang and Yuzhi Wang and Zhaoji Wang and Zhengtao Wang and Zhengtao Wang and Zhexu Wang and Chu Wei and Qianqian Wei and Haoning Wu and Wenhao Wu and Xingzhe Wu and Yuxin Wu and Chenjun Xiao and Jin Xie and Xiaotong Xie and Weimin Xiong and Boyu Xu and Jinjing Xu and L. H. Xu and Lin Xu and Suting Xu and Weixin Xu and Xinran Xu and Yangchuan Xu and Ziyao Xu and Jing Xu and Jing Xu and Junjie Yan and Yuzi Yan and Hao Yang and Xiaofei Yang and Yi Yang and Ying Yang and Zhen Yang and Zhilin Yang and Zonghan Yang and Haotian Yao and Xingcheng Yao and Wenjie Ye and Zhuorui Ye and Bohong Yin and Longhui Yu and Enming Yuan and Hongbang Yuan and Mengjie Yuan and Siyu Yuan and Haobing Zhan and Dehao Zhang and Hao Zhang and Wanlu Zhang and Xiaobin Zhang and Yadong Zhang and Yangkun Zhang and Yichi Zhang and Yizhi Zhang and Yongting Zhang and Yu Zhang and Yutao Zhang and Yutong Zhang and Zheng Zhang and Haotian Zhao and Yikai Zhao and Zijia Zhao and Huabin Zheng and Shaojie Zheng and Longguang Zhong and Jianren Zhou and Xinyu Zhou and Zaida Zhou and Jinguo Zhu and Zhen Zhu and Weiyu Zhuang and Xinxing Zu},
      year={2026},
      eprint={2507.20534},
      archivePrefix={arXiv},
      primaryClass={cs.LG},
      url={https://arxiv.org/abs/2507.20534}, 
}

\appendix


\section{Limitations}
\label{sec:limitations}

MCP-Atlas reflects a specific set of design choices, and several factors bound the conclusions that can be drawn from the results.

\paragraph{Judge-model ceiling.}
All evaluation is performed by large language models. We mitigate this with three independent judges but absolute pass rates should be read modulo a 2--5 point cross-judge band. Changes in judge model version between paper releases and follow-up evaluations may also shift reported numbers, so we recommend always reporting the judge used and pinning the judge version.

\paragraph{Snapshot of a moving ecosystem.}
Servers, tools, and underlying APIs are live. The current leaderboard reflects a May 2026 snapshot of server behavior. Rate-limit policies, authentication schemes, and schema versions change without notice, and a small number of tasks may become unsolvable or trivially solvable over time. We pin server container versions and will publish re-scoring runs when upstream drift is detected, but reproducibility at the level of individual task trajectories is bounded by upstream stability.

\paragraph{English-only prompts.}
All 1{,}000 tasks are authored in English, targeting English-language server responses. Non-English tool use, script handling, and localization failures are outside the current scope.

\paragraph{Provider-specific harness defaults.}
Where a provider recommends a native tool-calling strategy (e.g., anthropic-native, openai-responses-api), we use it. This follows published best practices but means that the harness is not fully identical across providers.

\paragraph{Tool-call budget.}
We fix a 100-call budget per task. A small number of failures stem from this ceiling rather than from model capability, particularly on long-horizon tasks that require retry-heavy recovery from real tool errors. Relaxing the budget is expected to shift absolute numbers upward but has not been shown to change relative rankings in internal tests.

\paragraph{Claim-extraction error band.}
Atomic claims were produced by a combination of expert authoring and automated candidate generation followed by manual review. Residual noise in the claim list produces noise on the achievable pass rate even for a perfect agent, as some claims are recoverable only under ambiguous interpretations of the prompt.

\paragraph{Reference Trajectory in Diagnostics.}
Because automated diagnostics use a verified reference trajectory, failure labels may reflect the author-provided evidence path when multiple very different evidence paths exist. We mitigate this by using claim-based pass/fail scoring, applying diagnostics only after failure, and instructing the diagnostic judge to treat the trajectory as one sufficient evidence trace rather than a canonical solution.

\paragraph{Broader impacts and risks.}
MCP-Atlas is intended to improve transparent evaluation of tool-using language-model agents by exposing reliability gaps before such systems are deployed in real workflows. The benchmark can help researchers and practitioners identify failures in tool discovery, evidence gathering, synthesis, and stopping behavior, thereby supporting safer and more accountable agent development. At the same time, better evaluation may accelerate the development of agents that interact more effectively with external services, including in settings where misuse, unauthorized data access, privacy leakage, or violations of upstream terms of service are possible. Live MCP-based evaluation also inherits risks from upstream services, including API drift, changing licenses, and accidental exposure of sensitive outputs if users connect the harness to private accounts. We mitigate these risks by sandboxing server containers, restricting network egress, withholding the private split, excluding credentials and secrets from the release, documenting upstream licenses, and releasing derived claims and scores rather than raw upstream tool-output snapshots. MCP-Atlas should therefore be used as a diagnostic benchmark for reliability and failure analysis, not as a certification that an agent is safe for safety-critical, privacy-sensitive, or high-stakes deployment.

\section{Server and Tool Inventory}
\label{app:servers}

\textbf{Server and Tool breakdown.} MCP-Atlas uses 36 production MCP servers exposing 220 distinct tools across all tasks. Each server is a community or vendor MCP implementation, deployed in a pinned container and exercised against live endpoints. Table~\ref{tab:server-inventory} lists every server, the number of tasks that require it, and its environment bucket assignment. Per-server tool catalogs, container image digests, and version pins are released alongside the dataset.

\begin{table}[h]
\centering
\small
\caption{Full server inventory. ``Tasks'' is the number of benchmark tasks in which the server is required. Several tasks require multiple servers, so the column sums to more than 1{,}000.}
\label{tab:server-inventory}
\setlength{\tabcolsep}{5pt}
\begin{tabular}{@{}lrl@{\hspace{2em}}lrl@{}}
\toprule
\textbf{Server} & \textbf{Tasks} & \textbf{Bucket} & \textbf{Server} & \textbf{Tasks} & \textbf{Bucket} \\
\midrule
oxylabs & 155 & Basic & git & 71 & Coding \\
filesystem & 133 & Productivity & mcp-server-code-runner & 70 & Coding \\
exa & 108 & Basic & osm-mcp-server & 68 & Basic \\
wikipedia & 99 & Basic & open-library & 68 & Basic \\
mongodb & 98 & Analytics & arxiv & 68 & Productivity \\
airtable & 96 & Analytics & twelvedata & 67 & Financial \\
mcp-code-executor & 93 & Coding & github & 67 & Coding \\
ddg-search & 92 & Basic & met-museum & 66 & Basic \\
national-parks & 85 & Basic & calculator & 65 & Basic \\
brave-search & 83 & Basic & desktop-commander & 58 & Coding \\
cli-mcp-server & 78 & Coding & google-maps & 56 & Basic \\
alchemy & 76 & Financial & clinicaltrialsgov & 51 & Basic \\
weather-data & 76 & Basic & whois & 49 & Basic \\
notion & 75 & Productivity & fetch & 49 & Basic \\
lara-translate & 72 & Basic & slack & 48 & Productivity \\
e2b-server & 72 & Coding & google-workspace & 40 & Productivity \\
 & & & memory & 36 & Productivity \\
 & & & pubmed & 30 & Basic \\
 & & & weather & 26 & Basic \\
 & & & context7 & 4 & Coding \\
\bottomrule
\end{tabular}
\end{table}

\textbf{Bucket Shares and Target Mix.} The distribution of tasks across the environment buckets is as follows: \textsc{Basic} (32\%), \textsc{Analytics} (12\%), \textsc{Productivity} (22\%), \textsc{Financial} (12\%), and \textsc{Coding} (22\%). Representative servers and tools are detailed in Table~\ref{tab:server-inventory}. Server usage varies across the benchmark, with higher representation of general-purpose servers such as search and filesystem.

\textbf{Server Selection Rationale.} The 36 MCP servers included in MCP-Atlas were selected to provide broad coverage across commonly used domains and use cases. Selection criteria included: (1) availability of stable, well-documented MCP implementations; (2) diversity across functional categories (search, analytics, productivity, finance, coding); (3) presence of multiple related tools per server to enable meaningful distractor selection; and (4) applicability to realistic multi-step workflows. The goal was to create a representative sample of the MCP ecosystem that tests agents across a variety of practical scenarios.

\paragraph{Containerization.}
Each server is packaged as a Docker container with a sandboxed filesystem (mounted read-only except for explicitly writable work directories) and an allow-listed network policy restricting egress to the set of upstream APIs the server legitimately requires. Containers are restarted between tasks to avoid cross-task state leakage. Evaluation runs pin each container to a specific image digest so that a re-run against the same digest yields identical server behavior modulo genuine upstream API drift.


\section{Evaluator Prompt}
\label{app:evaluator-prompt}



Each ground-truth claim is scored independently by the evaluator judge. The judge receives the claim text and the model's final answer and is asked to assign one of three categorical labels: \emph{fulfilled} (mapped to 1.0), \emph{partially\_fulfilled} (0.5), or \emph{not\_fulfilled} (0.0), along with a structured-output justification and a confidence value. The full prompt template is reproduced verbatim below.

\begin{quote}\small\ttfamily
You are evaluating how well a model's response addresses a specific expert-defined claim.

SCORING CRITERIA:\\
- fulfilled: Claim is completely and accurately addressed. The response covers all key details.\\
- partially\_fulfilled: Claim is partially addressed. The response covers some but not all key details.\\
- not\_fulfilled: Claim is not addressed. The response does not include any key details.

NUMERICAL COMPARISON GUIDELINES:\\
- For numerical values, use reasonable approximation thresholds:\\
\hspace*{1em}* Exact match NOT required for decimals\\
\hspace*{1em}* Values within 5\% of the claimed number are considered matching\\
\hspace*{1em}* For percentages, $\pm$1 percentage points is acceptable\\
\hspace*{1em}* Round to appropriate significant figures based on context\\
- Consider the precision appropriate to the domain:\\
\hspace*{1em}* Scientific measurements may need higher precision\\
\hspace*{1em}* General statistics/estimates can have looser matching\\
\hspace*{1em}* Financial figures should match to reasonable business precision (e.g., millions/billions don't need exact cents)\\
- If a number is expressed differently but mathematically equivalent (e.g., ``0.5'' vs ``50\%'' vs ``half''), consider it a match

CLAIM TO EVALUATE:\\
\{claim\}

MODEL RESPONSE TO ANALYZE:\\
\{response\}

INSTRUCTIONS:\\
1. Determine if the core requirement of the claim is met in the response\\
2. Check if all key components from the claim appear substantively in the response\\
\hspace*{1em}- For numerical values, apply the flexible matching guidelines above\\
\hspace*{1em}- Focus on whether the same magnitude and meaning are conveyed\\
3. Assign the appropriate coverage\_outcome\\
4. Provide specific justification referencing what was/wasn't covered\\
\hspace*{1em}- When numbers differ slightly, note if they're within acceptable range\\
5. Provide a confidence level (0.0-1.0) for your assessment

Be rigorous but fair in your assessment. Focus on whether the response conveys the same information as the claim, not on exact numerical precision unless precision is critical to the claim's meaning.
\end{quote}

The judge is invoked once per claim, not once per task, so a 5-claim task issues 5 independent evaluator calls. Per-task coverage is then computed as the mean of the 5 claim scores (mapped from the categorical labels), and the task passes at coverage e.g. $\geq 0.75$ for the main analyses. Per-claim independence avoids loss-of-context issues that would arise if the judge had to score every claim against the same response in a single call, especially for the 4\% of tasks with more than 10 claims.

\subsection{Confidence Intervals}
\label{sec:ci}

We report 95\% confidence intervals on pass@0.75 using the nonparametric bootstrap~\citep{efron1979bootstrap}. For each model, we draw $10{,}000$ resamples of $N=1000$ tasks with replacement from the per-task coverage scores, recompute pass@0.75 on each resample, and take the 2.5th and 97.5th percentiles of the resulting distribution as the 95\% CI. Intervals are reported as half-widths, e.g.\ $78.2\% \pm 2.5\%$.


\section{Task Authoring Templates and Example Task}
\label{app:worked-example}




To maintain high data quality, we provided expert annotators with a comprehensive internal manual governing task design and verification. Due to the proprietary nature of these full instructions, we provide a summary of the core requirements below. On average, each task required \textbf{4.6 hours} of expert labor to author, verify, and document all required deliverables.

Task authors and reviewers were compensated under contractual terms that met or exceeded applicable local wage requirements. No unpaid volunteer labor was used. We do not release the full internal authoring manual because it contains proprietary operational procedures, access-control details, and private benchmark templates.

\subsection{Task Authoring Requirements}
Annotators were instructed to generate prompts that are:
\begin{itemize}
    \item \textbf{Tool-Dependent \& Stable:} Tasks must require external data to solve and rely on information that remains valid throughout the evaluation window.
    \item \textbf{Unambiguous \& Natural:} Prompts must avoid subjective language and use conversational phrasing that does not reveal internal system mechanics or the intended solution path.
    \item \textbf{Reasoning-Intensive:} Authors were encouraged to create multi-step requests requiring the model to resolve indirect references or filter datasets across multiple sources.
\end{itemize}

\subsection{Tool Environment Construction}
For each task, annotators curated a custom tool environment containing:
\begin{itemize}
    \item \textbf{Sufficiency:} A set of relevant tools that provide the minimum necessary evidence to solve the prompt.
    \item \textbf{Distractor Integration:} Plausible but irrelevant tools were included to test the model's selection logic and its ability to ignore domain-unrelated APIs.
    \item \textbf{Multi-Source Complexity:} For multi-source tasks, environments were designed to require the combination of evidence across different tool families without introducing unnecessary ambiguity.
\end{itemize}

\subsection{Model Evaluation and Failure Taxonomy}
During task authoring, reviewers used a coarse three-way checklist—tool selection, tool input, and interpretation—to catch obvious solvability and rubric issues. This authoring checklist is distinct from the final 11-mode automated diagnostic taxonomy used for post-hoc model failure analysis in Appendix~\ref{app:taxonomy}.

To ensure consistent human review, annotators classify every model failure according to a specific taxonomy:
\begin{enumerate}
    \item \textbf{Tool Selection Error:} Selecting inappropriate tools or failing to utilize a necessary API.
    \item \textbf{Tool Input Error:} Passing incorrect, incomplete, or poorly constructed parameters (e.g., wrong date, entity, or query filter).
    \item \textbf{Interpretation Error:} Obtaining the correct data but misreading, misranking, or miscalculating the returned output, or ignoring a key constraint.
\end{enumerate}

\subsection{Required Deliverables}
In addition to the environment setup, annotators produced a \textit{Reference Trajectory} (a gold-standard sequence of calls with rationales) and a \textit{Grounded Final Answer} (a precise response strictly supported by tool outputs).

\subsection{Example Task}
\label{sec:example_task}

To illustrate the task schema, consider the following example:

\paragraph{Prompt.}
\begin{quote}
\emph{``I'm researching papers on advertisement effectiveness and comparing it
to our own online database advertising data. There's a 2024 paper by Jane
Castleman that deals with ad control effectiveness, can you get me the abstract?
I believe it mentions ad locality, for which I will also need to ask you for the
date of our campaign with the biggest engagement rate, started during the
2015--2023 period, and its locality.''}
\end{quote}

\paragraph{Enabled tools.}
Tools marked with $\dagger$ are distractors.

\smallskip
\noindent\textbf{Task-relevant tools.}

\noindent
\begin{tabular}[t]{@{}l@{\qquad}l@{}}
\texttt{arxiv\_search\_papers} &
\texttt{notion\_API-post-search} \\
\texttt{notion\_API-post-database-query} &
\end{tabular}

\smallskip
\noindent\textbf{Distractor tools.}

\noindent
\begin{tabular}[t]{@{}l@{\qquad}l@{}}
\texttt{notion\_API-retrieve-a-database}$^\dagger$ &
\texttt{notion\_API-retrieve-a-block}$^\dagger$ \\
\texttt{notion\_API-retrieve-a-page}$^\dagger$ &
\texttt{fetch\_fetch}$^\dagger$ \\
\texttt{memory\_read\_graph}$^\dagger$ &
\texttt{memory\_search\_nodes}$^\dagger$ \\
\texttt{slack\_channels\_list}$^\dagger$ &
\texttt{slack\_conversations\_history}$^\dagger$ \\
\texttt{slack\_conversations\_replies}$^\dagger$ &
\texttt{slack\_conversations\_search\_messages}$^\dagger$ \\
\texttt{whois\_whois\_domain}$^\dagger$ &
\texttt{whois\_whois\_tld}$^\dagger$.
\end{tabular}

\paragraph{Reference trajectory.}
\begin{enumerate}
    \item
    \begin{tabular}[t]{@{}ll@{}}
    \textbf{Tool:} & \texttt{arxiv\_search\_papers} \\
    \textbf{Query:} & \emph{``jane castleman ad locality 2024''} \\
    \textbf{Result:} & retrieve the paper abstract.
    \end{tabular}

    \item
    \begin{tabular}[t]{@{}ll@{}}
    \textbf{Tool:} & \texttt{notion\_API-post-search} \\
    \textbf{Query:} & \emph{``advertising''} \\
    \textbf{Result:} & identify the relevant advertising database.
    \end{tabular}

    \item
    \begin{tabular}[t]{@{}ll@{}}
    \textbf{Tool:} & \texttt{notion\_API-post-database-query} \\
    \textbf{Argument:} &
    \texttt{database\_id = 21b97551-844e-8068-b387-fe7a56b04348} \\
    \textbf{Result:} & retrieve the campaign date.
    \end{tabular}
\end{enumerate}

\paragraph{Claims.}
\begingroup
\renewcommand{\labelenumi}{\textbf{C\arabic{enumi}.}}
\begin{enumerate}
    \item There is a 2024 paper by Jane Castleman titled
    \emph{``Why am I Still Seeing This: Measuring the Effectiveness Of Ad
    Controls and Explanations in AI-Mediated Ad Targeting Systems.''}

    \item The abstract of that paper is:
    \emph{``Recently, Meta has shifted towards AI-mediated ad targeting
    mechanisms [... abridged for paper].''}

    \item Three advertising campaigns tie for the highest engagement rate, each
    with an engagement rate of 15\%.

    \item The starting dates of the three winning advertising campaigns are
    2022-06-24, 2019-09-20, and 2017-09-09.

    \item The localities of the three winning advertising campaigns are
    \emph{National}, \emph{International}, and \emph{International}.
\end{enumerate}
\endgroup

\section{Failure Mode Taxonomy}
\label{app:taxonomy}
The cognitive failure-mode columns in Table~\ref{tab:tool-call-failures} and Table~\ref{tab:cognitive-failures} are short labels for richer behavioral categories. Table~\ref{tab:taxonomy-defs} gives the full definition of every failure mode, the short label used in the body, and a representative concrete example. The same definitions are used by the LLM judge during diagnosis. The taxonomy has 11 modes: 4 in the tool-call family and 7 in the cognitive family. Diagnosis scope. We run the diagnostic prompt only for task–model pairs whose claim coverage is below the main pass threshold, i.e. coverage < 0.75. The diagnostic labels are used only for post-hoc failure analysis and never for pass/fail scoring. 

\begin{table}[h]
\centering
\small
\caption{Failure-mode definitions and representative examples. The 4 tool-call modes correspond to the columns in Table~\ref{tab:tool-call-failures} and the 7 cognitive modes correspond to the columns in Table~\ref{tab:cognitive-failures}.}
\label{tab:taxonomy-defs}
\setlength{\tabcolsep}{4pt}
\renewcommand{\arraystretch}{1.25}
\begin{tabular}{@{} >{\raggedright\arraybackslash}p{0.23\textwidth} p{0.3\textwidth} p{0.45\textwidth} @{}}
\toprule
\textbf{Mode (label)} & \textbf{What it captures} & \textbf{Example} \\
\midrule
\multicolumn{3}{@{}l}{\textit{Tool-call family}} \\
\midrule
Malformed (\texttt{malformed\_call}) &
Right tool, wrong parameters: missing arguments, bad types, or wrong values. &
Called \texttt{mongodb\_find} with column name \texttt{revenue} when the schema field is \texttt{Revenue\_USD}. \\
Wrong Tool (\texttt{wrong\_tool}) &
Picked a tool that cannot answer the subtask, even though a correct tool was available. &
Used \texttt{wikipedia\_search} for a fact that lives only in the Airtable database. \\
No Tool Use (\texttt{no\_tool\_use}) &
Answered from internal knowledge without calling any tool, even though tools were required. &
Reported a historical date directly without querying any source. \\
Err. Recovery (\texttt{err\_recovery}) &
A tool returned an error and the model could not adapt: it retried identically, looped, or gave up. &
Retried the same call five times instead of backing off or trying a different server. \\
\midrule
\multicolumn{3}{@{}l}{\textit{Cognitive family}} \\
\midrule
Task Misund. (\texttt{task\_mis-} \texttt{understanding}) &
Answered a different question than the one asked, or missed a key requirement in the prompt. &
Prompt asked for ``average revenue in December''; the model returned total revenue for the full year. \\
Faulty Synth. (\texttt{faulty\_synthesis}) &
Had the right tool outputs but combined or interpreted them incorrectly. Not a logic error. &
Queried the right table, got the right rows, then averaged the wrong column in the final answer. \\
Misparsing (\texttt{response\_misparsing}) &
Got valid tool output but misread its structure or extracted the wrong field. &
Tool returned a list of 10 records but the model picked the wrong row or read the wrong field. \\
Early Term. (\texttt{early\_termination}) &
Understood the task but stopped before completing all the required steps. &
Found one half of a two-part answer and produced a final answer without addressing the other half. \\
Hallucinated (\texttt{hallucinated\_fact}) &
Stated something in the final answer that did not appear in any tool output. &
Tool returned a population of 45{,}000 but the model wrote 54{,}000 in its answer. \\
Logical Err. (\texttt{logical\_error}) &
Multi-step reasoning chain was flawed even though the underlying data were correct. &
Correctly retrieved the date and database records, then applied the wrong conditional to filter them. \\
Constraint (\texttt{constraint\_} \texttt{violation}) &
Ignored an explicit condition or filter stated in the prompt. &
Prompt said ``only premium units built in 2017'' but the model queried all units across all years. \\
\bottomrule
\end{tabular}
\end{table}

\subsection{Diagnosis Prompt template}

The diagnostic prompt includes the reference trajectory because post-hoc failure attribution requires knowing which tool-accessible evidence was sufficient to satisfy the missed claims. The trajectory is a verified evidence trace, not a unique target policy. The diagnostic judge is instructed to use it to localize failures relative to the missed claims—for example, whether the model failed to seek required evidence, queried an unsuitable source, misread a returned field, or stopped before collecting all required evidence. It is not used to penalize alternative valid plans, different tool orderings, extra calls, parallel calls, or different parameterizations that produce the same claim-supporting evidence.

Prompt:

\begin{quote}\small\ttfamily
You are diagnosing why a model failed on an MCP (Model Context Protocol) tool-use evaluation task.

The model was connected to live MCP tool servers in sandboxed Docker environments and had to use tools to answer a multi-step question. Its final text response was scored against ground-truth claims.

TASK CONTEXT:\\
Task ID: \{task\_id\}\\
Original Prompt: \{prompt\}\\
Coverage Score Achieved: \{coverage\_score:.2\%\}\\
\{limit\_context\}

=== STEP 1. ANCHOR ON WHAT SHOULD HAVE HAPPENED ===

EXPECTED BEHAVIOR (Ground Truth Trajectory):\\
\{traj\_text\}\{traj\_suffix\}

EXPECTED CLAIMS TO ADDRESS:\\
\{gtfa\_claims\}

=== STEP 2. EXAMINE WHAT THE MODEL ACTUALLY DID ===

MODEL'S ACTUAL BEHAVIOR (Enriched Agent Trajectory):\\
Each turn shows: assistant reasoning, tool calls with parameters/status/errors/output summaries, parallel call detection, and final answer.\\
Examine the trajectory turn-by-turn and assess PROGRESSION:\\
- Is the agent making forward progress, or looping on the same call?\\
- Does it retry identically after a tool error, or adapt parameters?\\
- Does it switch tools unproductively when a single correct tool was available?\\
- Does it stop while required data is still missing?\\
\{agent\_trace\}

MODEL'S FINAL TEXT RESPONSE (this is what was scored against the claims):\\
\{final\_response\_text\}

=== STEP 3. CONSIDER WHICH CLAIMS WERE MISSED AND WHY ===

MISSED OR PARTIAL CLAIMS (with scorer reasoning):\\
The scorer below has already evaluated each claim against the model's final text response. Use the scorer's reason for each missed/partial claim as direct evidence when picking the failure mode.

\{missed\_text\}

=== STEP 4. MAP TO A FAILURE MODE ===

FAILURE MODE DEFINITIONS:

TOOL CALL FAMILY -- Problems with how the model interacted with tools:\\
- malformed\_call: Right tool, wrong parameters: missing arguments, bad types, or wrong values. (e.g., Called mongodb\_find with column name 'revenue' when the schema field is 'Revenue\_USD'.)\\
- wrong\_tool: Picked a tool that cannot answer the subtask, even though a correct tool was available. (e.g., Used wikipedia\_search for a fact that lives only in the Airtable database.)\\
- no\_tool\_use: Answered from internal knowledge without calling any tool, even though tools were required. (e.g., Reported a historical date directly without querying any source.)\\
- err\_recovery: A tool returned an error and the model could not adapt: it retried identically, looped, or gave up. (e.g., Retried the same call five times instead of backing off or trying a different server.)

COGNITIVE FAMILY -- Problems with how the model reasoned about the task:\\
- task\_misunderstanding: Answered a different question than the one asked, or missed a key requirement in the prompt. (e.g., Prompt asked for `average revenue in December'; the model returned total revenue for the full year.)\\
- faulty\_synthesis: Had the right tool outputs but combined or interpreted them incorrectly. Not a logic error. (e.g., Queried the right table, got the right rows, then averaged the wrong column in the final answer.)\\
- response\_misparsing: Got valid tool output but misread its structure or extracted the wrong field. (e.g., Tool returned a list of 10 records but the model picked the wrong row or read the wrong field.)\\
- early\_termination: Understood the task but stopped before completing all the required steps. (e.g., Found one half of a two-part answer and produced a final answer without addressing the other half.)\\
- hallucinated\_fact: Stated something in the final answer that did not appear in any tool output. (e.g., Tool returned a population of 45,000 but the model wrote 54,000 in its answer.)\\
- logical\_error: Multi-step reasoning chain was flawed even though the underlying data were correct. (e.g., Correctly retrieved the date and database records, then applied the wrong conditional to filter them.)\\
- constraint\_violation: Ignored an explicit condition or filter stated in the prompt. (e.g., Prompt said `only premium units built in 2017' but the model queried all units across all years.)

JUDGE REASONING PROCESS (do this internally before selecting a mode):\\
(a) Locate the specific turn or sentence in the final response where the model went off-track.\\
(b) State what should have happened instead (using the ground-truth trajectory and claims as reference).\\
(c) Decide whether the failure is in tool interaction (Tool Call family) or reasoning/synthesis (Cognitive family).\\
(d) Pick the most specific mode from that family.

JUDGE RULES:\\
1. Pick the most specific mode that fits. Common tricky boundaries:\\
\hspace*{1em}- response\_misparsing vs faulty\_synthesis: misparsing = read the wrong field/row from a tool output; faulty\_synthesis = read everything right but combined/aggregated wrong.\\
\hspace*{1em}- logical\_error vs faulty\_synthesis: logical\_error = the multi-step reasoning chain itself was flawed (wrong conditional, wrong filter); faulty\_synthesis = combined correct outputs incorrectly without a clear logic error.\\
\hspace*{1em}- hallucinated\_fact vs no\_tool\_use: hallucinated\_fact = stated a fact not present in any tool output; no\_tool\_use = bypassed tools entirely for a fact that required a tool.\\
\hspace*{1em}- malformed\_call vs err\_recovery: malformed\_call = first incorrect parameter usage; err\_recovery = same error reproduced repeatedly without adapting.\\
2. Identify the PRIMARY failure: the single mode that best explains the coverage gap.\\
3. Identify ALL failures in the trajectory (primary + contributing). Secondary failures use the same 11-mode vocabulary.\\
4. For each failure, mark is\_root\_cause=true only if it caused other failures downstream. If there is only one failure, mark it root cause.\\
5. faulty\_synthesis should only be used when the error is NOT better explained by logical\_error, response\_misparsing, or hallucinated\_fact.\\
6. Confidence calibration: 0.9-1.0 when the trajectory and scorer evidence both clearly point to one mode; 0.6-0.8 when the mode is the best fit but a secondary mode is close; below 0.6 when the trajectory is ambiguous.\\
7. Write a 1-2 sentence summary stating what the model did wrong and why a claim was missed. Cite the specific turn or final-response sentence.

IMPORTANT: This task scored below 1.0, meaning at least one claim was not satisfied. You \textbf{MUST} select a failure mode from the 11 listed above.
\end{quote}


\section{Cross-Judge Agreement (Extended)}
\label{app:judge-comparison}

Table~\ref{tab:judge-extended} reports the pass rate at coverage $\geq 0.75$ for each of the 20 models under each of the three evaluator judges, along with the per-model range across judges and the rank under each judge. 

\begin{table}[h]
\centering
\small
\caption{Pass rate (\%) at coverage $\geq 0.75$ under each evaluator judge. Range is the difference between the highest and lowest of the three judges. Ranks are computed among the 20 models under each judge.}
\label{tab:judge-extended}
\setlength{\tabcolsep}{4pt}
\renewcommand{\arraystretch}{1.05}
\begin{tabular}{@{}lrrrrrrr@{}}
\toprule
\textbf{Model} & \textbf{Gemini-3.1-Pro} & \textbf{GPT-5.4} & \textbf{Opus 4.6} & \textbf{Range} & \textbf{$r_{\text{Gem}}$} & \textbf{$r_{\text{GPT}}$} & \textbf{$r_{\text{Opus}}$} \\
\midrule
Muse Spark                    & 82.2 & 79.1 & 79.6 & 3.1 &  1 &  1 &  1 \\
Claude Opus 4.7               & 79.1 & 75.1 & 74.5 & 4.6 &  2 &  3 &  5 \\
Gemini 3.1 Pro Preview        & 78.2 & 75.8 & 76.4 & 2.4 &  3 &  2 &  2 \\
Claude Opus 4.6               & 76.8 & 73.3 & 74.8 & 3.5 &  4 &  4 &  4 \\
GLM-5.1                       & 75.6 & 72.8 & 75.2 & 2.8 &  5 &  5 &  3 \\
GPT-5.5                       & 75.3 & 72.0 & 71.9 & 3.4 &  6 &  6 &  6 \\
GPT-5.4                       & 70.6 & 68.5 & 69.9 & 2.1 &  7 &  7 &  7 \\
Gemini 3 Pro Preview          & 70.3 & 68.0 & 68.8 & 2.3 &  8 &  8 &  8 \\
Claude Opus 4.5               & 69.8 & 66.3 & 68.5 & 3.5 &  9 & 10 &  9 \\
Claude Sonnet 4.6             & 69.5 & 66.8 & 67.4 & 2.7 & 10 &  9 & 10 \\
GPT-5.2                       & 67.6 & 65.2 & 66.2 & 2.4 & 11 & 11 & 11 \\
Kimi K2.5                     & 64.4 & 59.9 & 63.0 & 4.5 & 12 & 12 & 12 \\
Gemini 3 Flash Preview        & 62.0 & 57.4 & 59.4 & 4.6 & 13 & 13 & 13 \\
Claude Sonnet 4.5             & 59.5 & 56.6 & 59.0 & 2.9 & 14 & 14 & 14 \\
GLM-4.7                       & 58.1 & 54.6 & 55.7 & 3.5 & 15 & 16 & 17 \\
Gemini 3.1 Flash Lite Preview & 57.1 & 54.3 & 56.1 & 2.8 & 16 & 17 & 16 \\
GPT-5.4 Mini                  & 56.7 & 54.9 & 57.7 & 2.8 & 17 & 15 & 15 \\
GPT-5.1                       & 50.1 & 47.9 & 50.8 & 2.9 & 18 & 18 & 18 \\
o3 Pro                        & 44.5 & 41.3 & 43.6 & 3.2 & 19 & 19 & 19 \\
Claude Haiku 4.5              & 40.2 & 37.1 & 38.7 & 3.1 & 20 & 20 & 20 \\
\midrule
\textbf{Mean}                 & \textbf{65.4} & \textbf{62.3} & \textbf{63.9} & \textbf{3.2} & & & \\
\bottomrule
\end{tabular}
\end{table}

\section{Coverage-Threshold Sensitivity}
\label{app:threshold_sens}

The main text reports pass rate at the $0.75$ coverage threshold, consistent with the MCP-Atlas leaderboard convention. Because the claims rubric returns a real-valued per-task coverage score, the choice of threshold is a reporting decision rather than a scoring decision. Table~\ref{tab:threshold-sensitivity} reports mean coverage alongside pass rate at $0.50$, $0.75$, and $0.9$ for all 20 models.

Two observations support using a single headline threshold. First, the
Spearman rank correlation between $0.50$ and
$0.75$ across the 20 models is high
($\rho = 0.965$),
indicating that rankings are largely preserved across thresholds. Second,
the absolute gap between the two pass rates is narrow at the top of the
leaderboard (Muse Spark: $89.6 \to 82.2$, a $7.4$~pp drop) and widens
toward the tail (Claude Haiku 4.5: $55.8 \to 40.2$, a $15.6$~pp drop).
This is the expected behavior of a well-behaved coverage metric:
partial-credit achievement is easier to earn than satisfying the stricter
headline pass criterion, and the gap between the two grows as overall
capability falls.

\begin{table}[h]
\centering
\small
\caption{Pass rate under three coverage thresholds, along with mean coverage across the 1{,}000-task set. Rankings are largely preserved across thresholds.}
\label{tab:threshold-sensitivity}
\setlength{\tabcolsep}{6pt}
\renewcommand{\arraystretch}{1.05}
\begin{tabular}{@{}lrrrr@{}}
\toprule
\textbf{Model} & \textbf{Mean Coverage} & \textbf{Pass@0.50 (\%)} & \textbf{Pass@0.75 (\%)} & \textbf{Pass@0.90 (\%)} \\
\midrule
Muse Spark                    & 0.862 & 89.6 & 82.2 & 71.2 \\
Claude Opus 4.7               & 0.838 & 89.5 & 79.1 & 63.2 \\
Gemini 3.1 Pro Preview        & 0.839 & 90.6 & 78.2 & 61.8 \\
Claude Opus 4.6               & 0.827 & 88.6 & 76.8 & 61.1 \\
GLM-5.1                       & 0.818 & 88.5 & 75.6 & 59.0 \\
GPT-5.5                       & 0.821 & 90.2 & 75.3 & 56.4 \\
GPT-5.4                       & 0.785 & 87.2 & 70.6 & 48.9 \\
Gemini 3 Pro Preview          & 0.790 & 85.9 & 70.3 & 52.7 \\
Claude Opus 4.5               & 0.769 & 82.4 & 69.8 & 53.9 \\
Claude Sonnet 4.6             & 0.780 & 84.1 & 69.5 & 55.0 \\
GPT-5.2                       & 0.755 & 81.4 & 67.6 & 50.6 \\
Kimi K2.5                     & 0.725 & 78.6 & 64.4 & 47.9 \\
Gemini 3 Flash Preview        & 0.720 & 78.2 & 62.0 & 47.5 \\
Claude Sonnet 4.5             & 0.697 & 75.7 & 59.5 & 45.0 \\
GLM-4.7                       & 0.685 & 74.1 & 58.1 & 43.7 \\
Gemini 3.1 Flash Lite Preview & 0.686 & 75.1 & 57.1 & 40.2 \\
GPT-5.4 Mini                  & 0.682 & 76.8 & 56.7 & 36.4 \\
GPT-5.1                       & 0.615 & 66.6 & 50.1 & 35.3 \\
o3 Pro                        & 0.542 & 58.4 & 44.5 & 31.6 \\
Claude Haiku 4.5              & 0.517 & 55.8 & 40.2 & 27.1 \\
\bottomrule
\end{tabular}
\end{table}


\section{Access, Licensing, and Reproducibility}
\label{app:access}

The released public split contains one row per task with the following fields: task identifier, enabled tool list, natural-language prompt, grounded factual claims, and reference trajectory. The reference trajectory is a serialized gold evidence trace containing the tool calls, arguments, dependencies, and task-specific returned evidence sufficient to solve the task. It is included to support reproducibility, smoke tests, solvability audits, and post-evaluation diagnostics. It is not used as a trajectory-matching target for pass/fail scoring.

\paragraph{Dataset access.}
We provide the 500-task public split, the task schema, and the claims. The 500-task private split is not released. It is held back for contamination monitoring and for future leaderboard integrity as new models are evaluated.

\paragraph{Harness and code.}
We provide the MCP-Atlas evaluation harness, claims evaluator, scoring pipeline, dependency pins, reference client wrapper, and scripts to regenerate all figures and tables from the released model-response and score files. The repository includes a local smoke test that runs without vendor model credentials. The agent harness allows up to 5 retries per task while evaluating each model, and a summary of the tool error(s) is fed back into the model context for fairness in evaluation. The eval allows partial response grading if the model hits the 100 tool call limit on the task. Full live re-evaluation requires provider API keys and, for some upstream MCP servers, service-specific credentials. These requirements and the affected experiments are documented in the README.

\paragraph{Licensing.}
The dataset (task prompts, claims, trajectories, distractor selections) is released under the Creative Commons Attribution 4.0 license. The harness and claims evaluator code are released under the Apache 2.0 license. Third-party MCP server implementations retain their upstream licenses, and we include an inventory of upstream licenses in the repository. Tool outputs are governed by upstream API terms, we release only curated claim-audit snippets and derived scores, not complete raw snapshots.


\paragraph{Compute for reproducing the leaderboard.} The full 20-model run required
$20 \times 1{,}000 = 20{,}000$ trajectories. Because the judge is invoked per claim
(mean 4.7 claims/task), the primary leaderboard required roughly
$20{,}000 \times 4.7 \approx 94{,}000$ evaluator calls. The three-judge sensitivity study
required roughly $3 \times 94{,}000 \approx 282{,}000$ calls. At observed mean trajectory
times (27--194 seconds per task), a single-model pass against all 1,000 tasks ran on a single
workstation with 8--12 concurrent clients, consuming between 2 and 18 wall-clock hours.
The primary evaluator (Gemini 3.1 Pro Preview) ran at 8--12 concurrent requests and completed
a 1,000-task pass in approx. 2 hours per model.

\section{Model Reasoning Configuration}
\label{app:model-reasoning-config}

We provide the reasoning configurations and parameters used for all of the models in Table~\ref{tab:model-configs}.

\begin{table}[]
\centering
\small
\setlength{\tabcolsep}{6pt}
\renewcommand{\arraystretch}{1.15}
\begin{tabular}{@{}lll@{}}
\toprule
\textbf{Model} & \textbf{Model ID} & \textbf{Config} \\
\midrule
Claude Opus 4.7    & \texttt{claude-opus-4-7}             & adaptive, effort=max     \\
Claude Opus 4.6    & \texttt{claude-opus-4-6}             & adaptive, effort=max     \\
Claude Opus 4.5    & \texttt{claude-opus-4-5}             & effort=high, budget=10k  \\
Claude Sonnet 4.6  & \texttt{claude-sonnet-4-6}           & reasoning=high           \\
Claude Sonnet 4.5  & \texttt{claude-sonnet-4-5-20250929}  & effort=high, budget=10k  \\
Claude Haiku 4.5   & \texttt{claude-haiku-4-5-20251001}   & reasoning=None (default) \\
\midrule
Gemini 3.1 Pro Preview        & \texttt{gemini-3.1-pro-preview}        & thinking=high \\
Gemini 3 Pro Preview          & \texttt{gemini-3-pro-preview}          & thinking=high \\
Gemini 3 Flash Preview        & \texttt{gemini-3-flash-preview}        & thinking=high \\
Gemini 3.1 Flash Lite Preview & \texttt{gemini-3.1-flash-lite-preview} & thinking=high \\
\midrule
GPT-5.5      & \texttt{gpt-5.5}            & reasoning=xhigh \\
GPT-5.4      & \texttt{gpt-5.4}            & reasoning=xhigh \\
GPT-5.2      & \texttt{gpt-5.2-2025-12-11} & reasoning=xhigh \\
GPT-5.4 Mini & \texttt{gpt-5.4-mini}       & reasoning=xhigh \\
GPT-5.1      & \texttt{gpt-5.1-2025-11-13} & reasoning=high  \\
o3 Pro       & \texttt{o3-pro}             & reasoning=None (default) \\
\midrule
\multicolumn{3}{@{}l}{\textbf{Meta}} \\
Muse Spark & \texttt{muse\_spark} & reasoning=max \\
\midrule
GLM-5.1   & \texttt{fireworks\_ai/glm-5p1}   & reasoning=None (default) \\
GLM-4.7   & \texttt{fireworks\_ai/glm-4p7}   & reasoning=None (default) \\
Kimi K2.5 & \texttt{fireworks\_ai/kimi-k2p5} & reasoning=None (default) \\
\bottomrule
\end{tabular}
\caption{Reasoning configurations for evaluated models.}
\label{tab:model-configs}
\end{table}

\section{Formalization}
\label{sec:formalization}

We formalize MCP-Atlas as a single-turn, multi-call agentic evaluation over real MCP servers with controlled tool exposure, utilizing a reference trajectory for analysis and claims-based scoring. Our notation follows the standard agent-server view adopted in recent MCP evaluations.

\subsection{Entities and Notation}

\textbf{Servers and Tools.} Let $\mathcal{S}=\{s_{1},...,s_{K}\}$ be the set of MCP servers. Server $s_{i}$ exposes a finite set of typed tools $T_{i}=\{t_{i,1},...,t_{i,|T_{i}|}\}$ following the MCP interface. The universe of tools is $T=\bigcup_{i=1}^{K}T_{i}$. Tools are assumed to be disjoint across servers.

\textbf{Task.} A task is defined as a tuple
$$\tau=(G, C, T_{\text{expose}}, \pi^{*}, \mathcal{C}^{*}),$$
where: (a) $G$ is the goal specification; (b) $C$ is the natural-language context/constraints; (c) $T_{\text{expose}}\subseteq T$ is the allow-listed tool set presented to the agent; (d) $\pi^{*}$ is the reference trajectory, a minimal dependency-covering sequence of tool calls used for diagnostics; and (e) $\mathcal{C}^{*}=\{c_{1},...,c_{m}\}$ is the claims list, comprising disjoint, independently verifiable propositions that the correct answer must contain.

\textbf{Targets vs. Distractors.} We define $T_{\text{req}}\subseteq T_{\text{expose}}$ as the target tools required by $\pi^{*}$, and $T_{\text{dist}}=T_{\text{expose}}\setminus T_{\text{req}}$ as the distractors.

\subsection{Interaction Model and Traces}

\textbf{Tool Signatures.} Each tool $t\in T$ defines a typed relation
$$t: \text{Args}_{t} \rightarrow \text{Out}_{t} \cup \text{Err}_{t}$$
where $\text{Args}_{t}$ is a JSON-schema-typed domain. Outputs are either a well-typed value in $\text{Out}_{t}$ or an error token $e\in \text{Err}_{t}$ (e.g., schema violation, rate limit).

\textbf{Interaction Model.} Given a single-turn prompt $(G, C)$, the agent issues tool calls from $T_{\text{expose}}$. A call at step $k$ is $u_{k}=\langle t_{k},a_{k}\rangle$, and the environment returns $o_{k}\in \text{Out}_{t_{k}}\cup \text{Err}_{t_{k}}$. The execution trace is $x_{1:K}=((u_{1},o_{1}),...,(u_{K},o_{K}))$. The agent must finally emit a textual answer $\hat{A}$.

\textbf{Planning View.} This interaction induces a finite-horizon POMDP. Rewards are attached post-hoc via the claims-based scoring function.

\subsection{Claims-Based Scoring}

\textbf{Answer-Centric Ground Truth.} Let $\mathcal{Y}$ be the space of model answers. The benchmark defines a scoring functional
$$S:\mathcal{Y}\times 2^{\mathcal{C}^{*}}\rightarrow[0,1],$$
which computes a rubricized correctness score by checking $\hat{A}\in\mathcal{Y}$ against the claims list $\mathcal{C}^{*}$. Pass/fail is determined solely by $S(\hat{A},\mathcal{C}^{*})$, independent of the trajectory $\pi^{*}$.



\end{document}